\documentclass[aps,twocolumn,nofootinbib,preprintnumbers,citeautoscript]{revtex4}
\usepackage{amsmath,amssymb,epsfig}
\usepackage{enumerate}
\usepackage{mathpazo}
\usepackage{braket}
\usepackage{color}
\usepackage[colorlinks=true,citecolor=blue,urlcolor=blue]{hyperref}
\newcommand{\ketbra}[2]{|#1\rangle \langle #2|}
\newcommand{\etal}{{\it{et al. }}}
\newcommand{\tr}{\operatorname{Tr}}
\newcommand{\be}{\begin{equation}}
\newcommand{\ee}{\end{equation}}
\newcommand{\ba}{\begin{eqnarray}}
\newcommand{\ea}{\end{eqnarray}}
\newcommand{\blu}{\color{blue}}
\newcommand{\grn}{\color{green}}
\newcommand{\red}{\color{red}}
\newcommand{\bla}{\color{black}}

\renewcommand\refname{References and Notes}

\def\K{{\mathcal K}}
\def\G{{\mathcal G}}
\def\be{\begin{equation}}
\def\ee{\end{equation}}
\def\lp{\ell_P}
\def\R{{\mathcal R}}
\def\R{{\mathcal R}}
\def\Q{{\mathcal Q}}
\def\N{{\mathcal N}}
\def\M{{\mathcal M}}
\def\W{{\mathcal W}}
\def\L{{\mathcal L}}
\def\H{{\mathcal H}}
\def\L{{\mathcal L}}
\def\K{{\mathcal K}}
\def\be{\begin{equation}}
\def\ee{\end{equation}}
\def\lp{\ell_P}
\def\deth{{\rm det~}h}
\def\a{\alpha}
\def\b{\beta}
\def\g{\gamma}
\def\d{\delta}
\def\e{\epsilon}
\def\f{\phi}
\def\fin{f_\infty}
\def\r{\rho}
\def\l{\lambda}
\def\k{\kappa}\def\m{\mu}\def\n{\nu}\def\s{\sigma}\def\l{\lambda}
\def\bnabla{\bar\nabla}
\def\bN{\bar{\N}}\def\bg{\bar{g}}
\def\bbox{\overline{\Box}}\def\beq{\begin{eqnarray}}\def\eeq{\end{eqnarray}}

\usepackage{graphicx}

\begin{document}

\title{Pearson Correlation Coefficient as a measure for Certifying and Quantifying High Dimensional Entanglement}

\author{C. Jebarathinam$^{1,\dagger}$, Dipankar Home$^{2}$, Urbasi Sinha$^{3}$}
\email{usinha@rri.res.in}
\affiliation{$^1$S. N. Bose National Centre for Basic Sciences, Block JD, Sector III, Salt Lake, Kolkata 700 106, India}
\altaffiliation{Department of Physics and Center for Quantum Frontiers of Research \& Technology (QFort), National Cheng Kung University, Tainan 701, Taiwan}
\affiliation{$^2$Centre for Astroparticle Physics and Space Science (CAPSS), Bose Institute, Block EN, Sector V, Salt Lake, Kolkata 700 091, India}
\affiliation{$^3$Light and Matter Physics, Raman Research Institute, Bengaluru-560080, India}


\begin{abstract}
 A  scheme for characterizing entanglement using the statistical measure of correlation given by the Pearson correlation coefficient (PCC) was recently  suggested
that has remained unexplored beyond the qubit case. Towards the application of this scheme for the high dimensional states, a key step has been taken in a very recent work by experimentally determining PCC and analytically relating it to Negativity for quantifying entanglement of the empirically produced bipartite pure state of spatially correlated photonic qutrits. Motivated by this work, we present here a comprehensive study of the efficacy of such an entanglement characterizing scheme for a range of bipartite qutrit states by considering suitable combinations of PCCs based on a limited number of measurements. For this purpose, we investigate the issue of necessary and sufficient certification together with quantification of entanglement for the two-qutrit states comprising maximally entangled state mixed with white noise and coloured  noise in two different forms respectively. Further, by considering these classes of states for $d=4$ and $5$, extension of this PCC based approach for higher dimensions ($d$) is  discussed.
\end{abstract} 

\pacs{}

\maketitle

\section{Introduction}

Seminal discoveries of the applications of quantum entanglement in cryptography \cite{ek}, superdense coding 
\cite{ben} and teleportation \cite{benn} have given rise to a rich body of works that have demonstrated the remarkable power of entanglement as resource for quantum communication and information processing tasks, ranging  from secure key distribution \cite{ABG+07}, quantum computational speed-up \cite{JL03}, reduction of communication complexity \cite{bruk,BCM+10}, to device-independent certification of genuine randomness \cite{PAM+10, NPS14}. These explorations have primarily focused on considering the two-dimensional (qubit) systems. Alongside, though, it is important to note that there have been a number of studies indicating a range of advantages gained by using high dimensional entangled states, for example, achieving  more robust quantum key distribution protocols with higher  key rate \cite{bech,CBK02,BCE+03,SV10}, ensuring increased security of the device independent key distribution protocols against even tiny imperfection in randomness generation \cite{hub}, enhancing quantum communication channel capacity \cite{bennett,WDF+05}, as well as lowering the rate of entanglement decay arising from atmospheric turbulence in the context of free-space quantum communication \cite{brun} and reducing the critical detection efficiency required for more robust tests of quantum nonlocality \cite{ver}.

Thus, in light of this promising potentiality  of high dimensional entangled states, the characterization of such experimentally produced entangled states is of much significance. Here it needs to be noted that the tomographic characterization of quantum states is constrained by the requirement to determine a large number of independent parameters depending upon the dimension of the system \cite{IOA17}. Hence, in order to obviate this difficulty, the study of characterization of high-dimensional entangled states based on a limited number of measurements has been attracting an increasing attention.  Further,  since which of the proposed schemes for characterizing entanglement would be most readily amenable to experimental implementation is a priori an open question,  the search for various effective schemes on this issue acquires considerable significance.
 On the one hand, there are schemes making use of entanglement witnesses to provide lower bounds on the entanglement measures \cite{SSV13,SSV14}, on the other hand, operational quantification of entanglement in a measurement-device-independent way has been analyzed within the context of a subclass of semiquantum nonlocal games \cite{SHR17}  and  this approach has been used \cite{SSC17} to provide measurement-device-independent bounds on entanglement quantifiers like Negativity.    Also, of particular interest in this context are the recent studies \cite{TDC+17,MGT+17,BVK+17} formulating approaches to provide sufficient characterization of bipartite high-dimensional entanglement based on determining a  lower bound to the entanglement of formation from a limited number of measurements. Among these approaches, the scheme used by Bavaresco \etal \cite{BVK+17} gives an optimal estimate of the lower bound for entanglement of formation,
and this scheme is easier to experimentally implement
because it involves only two local measurements in each wing of the bipartite system. 
 A different approach \cite{SH18} based on the violation of entropic  inequalities
witnessing steerability of high dimensional entanglement with only two local measurements, too, has been shown to provide   an optimal lower bound to 
the entanglement of formation. 

However,  all such approaches focusing essentially on providing bounds on entanglement measures,  do not provide quantification of entanglement in terms of determining the actual value of an entanglement measure like entanglement of formation or Negativity. On the other hand,
while the characterization of entanglement for bipartite and multipartite qubit states was earlier discussed in terms of appropriate inequalities involving Bell correlations \cite{Roy05}, a recent relevant study \cite{DAC17} proposes using the Son-Lee-Kim (SLK) inequality (a bipartite Bell-type inequality whose violation can show nonlocality of high-dimensional states) for entanglement
characterization by relating the nonzero value of the measurable SLK function to Negativity (concurrence) in the case of high-dimensional pure states (isotropic mixed states) based on measurements of an appropriately chosen set of observables. However, this approach has the limitation that  nonzero value of the SLK function is not a sufficient condition for certifying  entanglement since there are separable mixed states for which the SLK function is nonzero for the measurements of the observables specified in this approach. Now, while such approaches make use of linear inequalities, there have also been studies \cite{GL06,AGL12} formulating nonlinear entanglement witnesses that are more effective in detecting entanglement than the linear entanglement witnesses; however, still not quantifying entanglement in the sense mentioned earlier.

Next, considering the other approaches that have  been proposed for the characterization of entanglement for high-dimensional bipartite systems,  the following 
are particularly noteworthy. A scheme based on the sum of mutual information using two mutually unbiased bases (MUBs)  has been invoked to certify various noisy mixed entangled states in higher-dimensional cases using the notion that a bipartite multidimensional state in even dimension can be regarded as an ensemble of bipartite qubit states \cite{HMK+16}; however, this scheme provides only sufficient criterion for detecting entanglement and  quantifies entanglement in terms of entanglement of formation, essentially  restricted to the maximally entangled state \cite{MBM15}. Another approach based on the notion of mutual predictability has led to the argument that the condition of the sum of  mutual predictabilities pertaining to MUBs exceeding a certain bound can serve as a necessary and sufficient criterion for certifying entanglement of pure and isotropic mixed states in any dimension  \cite{SHB+12}. On the other hand, using measurements pertaining to correlations present in two appropriately chosen MUBs, the experimental feasibility of a scheme \cite{EKH17} has been argued that can determine essentially a lower bound to the entanglement of formation for any state, while providing only sufficient certification  of entanglement of the coloured-noise and isotropic mixed states. 

The preceding discussion, thus, underscores the lack of schemes that, apart from necessary and sufficient certification,  can also quantify high dimensional entanglement in the sense of  determining the actual value of an appropriate entanglement measure in terms of a limited number of experimentally measurable quantities.
Of course, in such analyses, it is assumed at the outset  that the empirical procedure for preparing a bipartite correlated state can specify 
it to be pure or mixed, and if mixed, the type of noise that is involved in the preparation procedure. 
The approach we adopt here is based on analytically linking an empirically accessible statistical measure of correlation
with a suitable  entanglement measure. For this purpose, Maccone et al. \cite{MBM15} had suggested the use of Pearson correlation coefficient \cite{pea} for entanglement characterization. The Pearson correlation coefficient (PCC) for any two random variables $A$ and $B$ is defined as

 \begin{equation} \label{PCCdef}
 \mathcal{C}_{AB}\equiv 
 \frac{\braket{AB}-\braket{A}\braket{B}}
 {\sqrt{\braket{A^2}-\braket{A}^2}\sqrt{\braket{B^2}-\braket{B}^2}},
 \end{equation}
whose values can lie between $-1$ and $1$, and  $\braket{\cdot}$ is an average value.
Note that although PCC is a well known measure of correlation that has been applied extensively in different areas of statistical applications, surprisingly, it has  so far been used in physics only in a few cases such as for quantifying the temporal correlation between classical trajectories in the context of synchronization problems \cite{BKO02}, for the quantification of synchronization in the context of temporal dynamics of local  observables of a bipartite quantum system \cite{BGP+17}, and for formulating Bell-CHSH type inequality in terms of PCCs \cite{PHB+17}.

Now, let us explore the application of PCC in the context of the following scenario: suppose a bipartite pure or mixed state is shared between Alice and Bob in an arbitrary dimension; Alice (Bob) performs two dichotomic measurements $A_1$ ($B_1$) and $A_2$($B_2$) on her (his) subsystem. Then, for $A_1=B_1=\sum_j a_j \ketbra{a_j}{a_j}$ and $A_2=B_2=\sum_j b_j \ketbra{b_j}{b_j}$, where $\{\ket{a_j}\}$ is mutually unbiased to $\{\ket{b_j}\}$, the following condition has been conjectured by Maccone et al. to certify entanglement of bipartite systems, i.e.,
 \begin{equation}
|\mathcal{C}_{A_1B_1}|+|\mathcal{C}_{A_2B_2}|>1, \label{s2pcc}
 \end{equation}
 is postulated to imply entanglement. However, this procedure based on PCCs has been applied for entanglement characterization restricted to \textit{only} the qubits \cite{MBM15}.

In this context, it is important to take note of the line of studies that has been recently initiated by measuring PCCs for a bipartite photonic qutrit pure state which has been produced using a novel pump beam modulation based technique \cite{GJK+17}. Subsequently, very recently, by analytically relating the experimentally measurable quantity PCC with Negativity as a measure of entanglement, the value of Negativity for the empirically prepared nearly maximally entangled state has been inferred, thereby constituting the first work using PCC demonstrating entanglement detection and quantification beyond the two qubit case \cite{Sinha2019}. While in that work, specifically, pure two qutrit states have been considered, in this paper we embark on a comprehensive study of the application of PCC based entanglement characterizing scheme. In particular, we explore the above mentioned conjecture of Maconne et al. by considering a range of mixed states like isotropic and two-types of coloured-noise mixed states, as well as the Werner and Werner-Popescu states in terms of the sum of suitable number of PCCs.

 Here it is relevant to note that the particular significance of the qutrit systems stems from the considerable practical advantages as compared to qubits that have been decisively shown in the context of quantum cryptography \cite{DCG+03}, quantum computation \cite{green}, and robustness against entanglement decay \cite{brun}; moreover, because of the intriguing nature of the relationship that has been pointed out for the qutrits between the magnitude of violation of Bell-type inequality and the amount of entanglement \cite{col,KKC+02,acin}, the study of entangled qutrits acquires an added fundamental significance.

A salient feature of our treatment worth stressing is that 
it is the idea of Negativity as a measure of entanglement that turns out to be useful for relating it to PCCs in a way 
that enables effective characterization of entanglement for the classes of states considered in this paper. Here it is relevant to recall that introduction of the idea of Negativity 
by Zyczkowski \etal \cite{ZHS+98} stimulated its use as an entanglement measure through demonstration  that it is an entanglement monotone for any finite-dimensional bipartite entangled state \cite{VW02}. Later, applications of this quantity, defining it as the absolute value of the sum of negative eigenvalues of partial transposed density matrix, were pointed out in different contexts like relating its lower bound to the violations of Bell-CHSH inequality  and steering inequality respectively \cite{MBL+13,Pus13}.
 A physical meaning of Negativity has been provided by arguing that Negativity can be viewed as an estimator of the  number of degrees of freedom of the  two subsystems that are entangled, as well as can  be viewed as determining in a device-independent way the minimum number of dimensions that contribute to the quantum correlation  \cite{ES13}. In this context, the relationship between Negativity and PCCs found in this paper can have interesting implications revealing further aspects of the physical meaning of Negativity for higher dimensional systems.

Now, let us summarize  the salient results obtained in  Section \ref{Two-qutrit} for the \textit{qutrit} case:

{\bf (a)} We consider maximally entangled state mixed with white noise in two different forms, \textit{isotropic mixed states} \cite{HH99,TV00,RC03} and \textit{Werner-Popescu states} \cite{HH99,Pop94}. For both these classes of mixed states, it is found that by appropriately choosing four mutually noncommuting bases which are \textit{not} MUBs, the sum of four PCCs being greater than $1$ provides the \textit{necessary} and \textit{sufficient condition} for certifying entanglement, as well as  the \textit{quantification} of entanglement is obtained  through an analytically derived \textit{monotonic relation} in terms of \textit{Negativity}. 

{\bf (b)}  We consider two types of \textit{coloured-noise} mixed with maximally entangled state.
In one of the types, 
coloured-noise state having perfect correlation in the computational basis is mixed with the maximally entangled state \cite{HMK+16}. For this family of states, we find that one can choose two appropriate MUBs  so that the sum of two PCCs being greater than $1$ gives the \textit{necessary} and \textit{sufficient condition} for certifying entanglement; \textit{quantification} of entanglement is also obtained similar to the earlier  cases in terms of  Negativity.

In the other type, coloured-noise state having anti-correlation in the computational basis is mixed with the maximally entangled state \cite{ETS15}.
For this class of states, we find that for the appropriately chosen four mutually noncommuting bases which are not MUBs, the sum of four PCCs being greater than $1$ furnishes the \textit{certification and quantification} of 
 entanglement, provided \textit{Negativity is nonvanishing}.

{\bf (c)} Considering the entanglement characterization of \textit{Werner state} \cite{Wer89} which, in any arbitrary dimension, is a mixture of projectors onto the antisymmetric subspace and white noise
in the higher dimensional case, it turns out that by using the sets of four appropriate mutually noncommuting bases, MUBs as well as non-MUBs, we can show the sum of four PCCs to be providing  \textit{sufficient criterion} for the certification of entanglement, as well as the \textit{quantification} of entanglement can be achieved  by relating it to Negativity. 

It is thus evident that for the effective characterization of entanglement using PCCs for the different types of qutrit mixed states, the number of measurements suffice to be limited to either only two or four MUBs or noncommuting bases. An interesting point to note is that while the schemes for efficient tomography and those invoking the notions of mutual information and mutual predictability  usually use MUBs, the approach proposed for entanglement characterization in terms of PCCs can work for some specific classes of states  like isotropic mixed states, a type of coloured-noise, Werner and Werner-Popescu states, even using mutually noncommuting bases that are not MUBs. This is similar to the case of nonlocality studies using Bell-type inequalities involving measurements pertaining to mutually noncommuting bases which do not necessarily need to be MUBs \cite{acin}. Here we may also mention that apart from its other applications, the procedure of entanglement characterization and quantification using PCCs in the qutrit case, together with the results of studies on the nonlocality of bipartite qutrit states 
can provide a powerful experimental platform for a comprehensive probing of hitherto unexplored quantitative aspects of  the relationship between entanglement and 
nonlocality \cite{col,KKC+02,acin,AGG05,BGS05,ZG08,JP11,BBM+14,DDG+17}.

\begin{figure}[!t]
\includegraphics[scale=0.49]{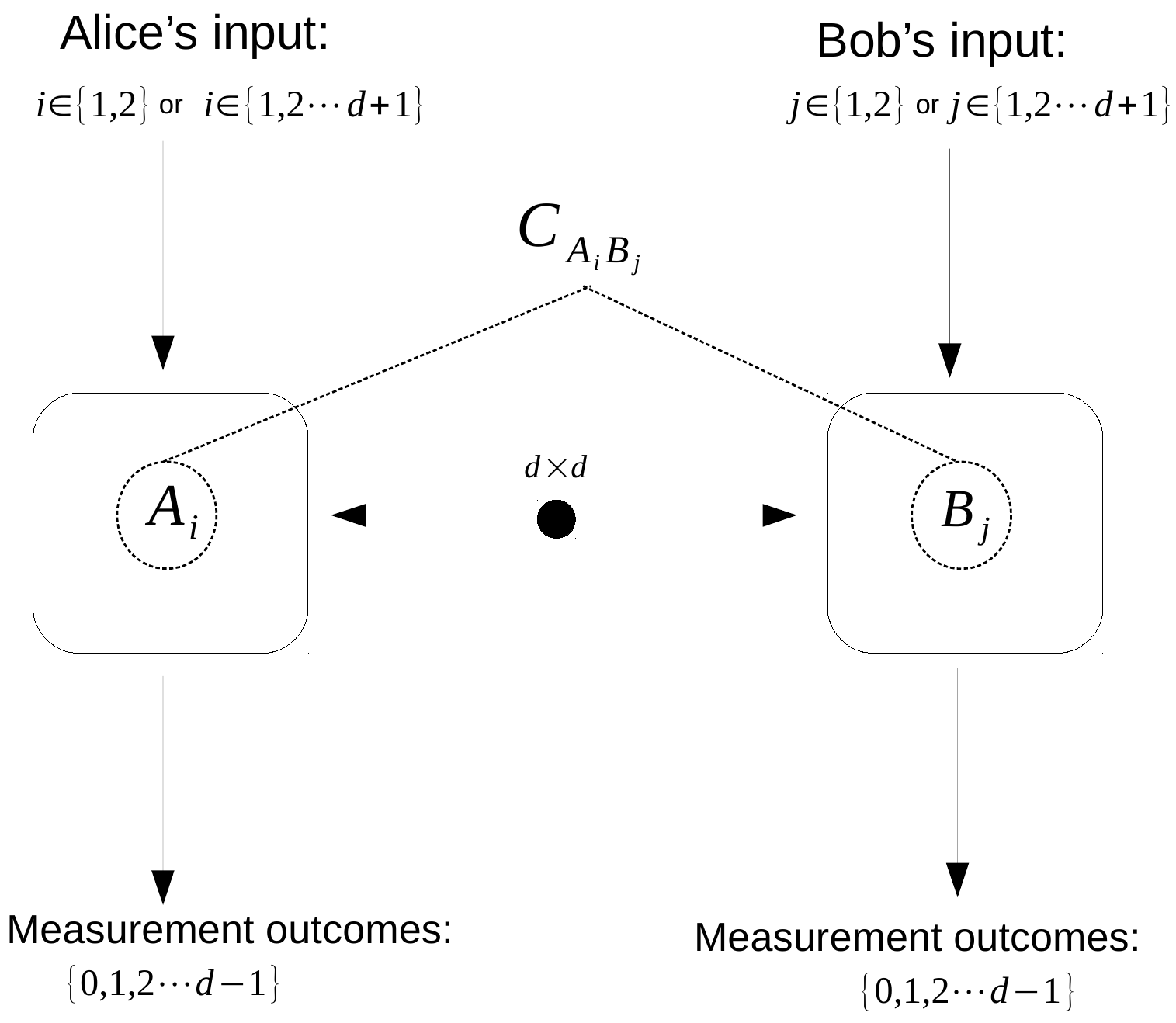}
\caption{\footnotesize Entanglement characterization approach based on the sum of Pearson correlation coefficients (PCCs). 
Two experimentalists, Alice and Bob, have access to the 
subsystems of a bipartite $d \times d$ quantum system. Alice and Bob perform two or $d+1$ local measurements in mutually unbiased bases or noncommuting bases. From the measurement statistics, Alice and Bob can
check whether the sum of two PCCs given by $C_{A_1B_1}+C_{A_2B_2}$ (in the case of pure states) or the sum of $d+1$ PCCs 
given by $\sum^{d+1}_{i,j=1} C_{A_iB_j}$ (in the case of mixed states) is greater than $1$ to determine whether the given bipartite quantum state is entangled or not.} 
\label{PCCfig}
\end{figure}

In Section III, towards exploring the potentiality of this method for higher dimensions $d > 3$, the results of   studies probing extension of this scheme for the dimensions $d=4$ and $5$ will be discussed, in particular, for the pure as well as  the isotropic, two types of coloured-noise, Werner  and Werner-Popescu mixed states
(see Fig. \ref{PCCfig} which gives a schematic outline of this entanglement characterization approach). 
We now proceed to delve into  the specifics, beginning with the case of isotropic mixed states.

\section{Two-qutrit states}\label{Two-qutrit}

\subsection{Isotropic mixed states}

Let us begin by writing the  general expression for  the two-qudit isotropic mixed state \cite{HH99,TV00,RC03} given by
\begin{equation}{\label{Isomixedsattedef}}
	\rho_I(F) 
	= \dfrac{1-F}{d^2 - 1} \, (\, \mathbb{I}- \ketbra{\phi_d^+}{\phi_d^+}\,) + \, F \, \ketbra{\phi_d^+}{\phi_d^+}
	\end{equation}
where $F = \braket{\phi_d^+|F|\phi_d^+}$ satisfying $0 \le  F \le 1$ is the fidelity of $\rho_I(F)$ and 
\begin{equation}
\ket{\phi_d^+}=\frac{1}{\sqrt{d}}\sum^{2}_{i=0} \ket{i} \otimes \ket{i}
\end{equation}
which is the maximally entangled state in dimension $d$ and
$\mathbb{I}$ is the identity matrix of dimension $d \times  d$. 
For the two-qudit isotropic mixed state $\rho_I(p)$, Negativity as defined in Ref. \cite{VW02} can be computed from the partial transposed density matrix
and is given by 
\begin{align}\label{nqudiso}
\mathcal{N}(\rho_I(F))=\max \Bigg\{ \dfrac{dF -1}{2} , 0 \Bigg \}		
\end{align}
which is nonzero if and only if $F>1/d$. Interestingly, it turns out that the two-qudit isotropic mixed state $\rho_I(F)$ is entangled
if and only if the same condition is satisfied, viz., $F>1/d$ \cite{HH99}. Therefore, it follows that the Negativity of this class of states as given by Eq. (\ref{nqudiso}) provides the necessary and sufficient
quantification of entanglement for any $d$.


For our purpose here for the necessary as well as sufficient certification of entanglement, we now construct the following set of four noncommuting bases which are not MUBs:
\begin{align}\label{csmub3}
\{\ket{a_j}\}=&\{\ket{0}, \ket{1}, \ket{2} \} \nonumber \\
 \{\ket{b_j}\}=&\{(\ket{0}+ \ket{1}+\ket{2})/\sqrt{3}, \nonumber \\
                &(\ket{0}+\omega \ket{1}+\omega^2\ket{2})/\sqrt{3} , \nonumber \\
                 &(\ket{0}+ \omega^2 \ket{1}+ \omega\ket{2})/\sqrt{3}\}, \nonumber \\
\{\ket{e_j}\}=&\{(\ket{0}+e^{i\pi/3} \ket{1}+e^{2i\pi/3}\ket{2})/\sqrt{3}, \nonumber \\
                & (\ket{0}- \ket{1}+\ket{2})/\sqrt{3}, \nonumber \\
                 &(\ket{0}+\omega^2 e^{i\pi/3} \ket{1}+\omega e^{2i\pi/3}\ket{2})/\sqrt{3}\}, \nonumber \\
\{\ket{g_j}\}=&\{(\omega^2\ket{0}+\omega \ket{1}-\ket{2})/\sqrt{3}, \nonumber \\
                &(\ket{0}+ \ket{1}-\ket{2})/\sqrt{3} , \nonumber \\
                 &(\omega \ket{0}+ \omega^2 \ket{1}- \ket{2})/\sqrt{3}\},               
\end{align}
where $\omega=e^{2i\pi/3}$. 
Here, the eigenvalues  $a_j$ of the computational basis \cite{NC00} are given by $a_0=+1$, $a_1=0$ and $a_2=-1$, the second basis
$\{\ket{b_j}\}$ corresponds to what we call the 
generalized $\sigma_x$-basis (with 
the eigenvalues $b_{0} = 0$, $b_{1}=\pm 1$, $b_{2}=\mp 1$), the third basis $\{\ket{e_j}\}$ corresponds to what we call the generalized $\sigma_y$-basis (with the
eigenvalues $b_{0} = +1$, $b_{1}=0$, $b_{2}=-1$)
and the eigenvalues  $g_j$ of the fourth basis are given by $g_0=+1$, $g_1=0$ and $g_2=-1$.\\

Here we may remark that what we call the generalized $\hat{\sigma}_{x}$ and the generalized $\hat{\sigma}_{y}$ bases mentioned above which will be used later are obtained from the general expression for the $d$-dimensional basis invoked by Scarani et al.\cite{SGB+06} in the context of studies related to the CGLMP inequality; also, used in the treatment by Spengler et al. \cite{SHB+12}. This eigenbasis $\{\Psi_x(a)\}$ of a $d$-dimensional observable as invoked by these authors can be written in terms of the computational basis as follows: 
\be
\Psi_x(a)\equiv\sum^{d-1}_{k=0}\frac{e^{i(2\pi/d)ak}}{\sqrt{d}}(e^{ik\phi_x}\ket{k}). \label{GB}
\ee
where $a=0,1,2....(d-1)$ label the different eigenvectors.
 For $d\ge3$, we call the basis
$\{\Psi_x(a)\}$ with $\phi_x=0$ and $\phi_x=\pi/d$ the generalized $\sigma_x$ basis and the generalized $\sigma_y$ basis
respectively. This terminology is used in the sense that in the case of $d=2$, the above expression reduces to the eigenbases corresponding to $\sigma_x$ and $\sigma_y$ 
observables respectively.

Next, using the earlier mentioned bases given by Eq.(\ref{csmub3}), we find that the necessary and sufficient certification of entanglement for the two-qutrit isotropic states can be obtained 
in terms  of the sum of four PCCs $\sum^4_{i=1}|\mathcal{C}_{A_iB_i}|$, 
where $A_1=B_1=\sum_j a_j \ketbra{a_j}{a_j}$, $A_2=B_2=\sum_j b_j \ketbra{b_j}{b_j}$, 
$A_3=B_3=\sum_j e_j \ketbra{e_j}{e_j}$ and $A_4=B_4=\sum_j g_j \ketbra{g_j}{g_j}$, 
whence the sum of these four PCCs is given by
 \begin{align}\label{d3isoNS}
 \sum^4_{i=1}|\mathcal{C}_{A_iB_i}|
 &=\frac{|9F-1|}{2}>1 \quad \text{iff} \quad  F>1/3. 
 \end{align}
See Appendix \ref{AIII} for the derivation of the above expression for the sum of
four PCCs. Now, from Eqs. (\ref{nqudiso}) and (\ref{d3isoNS}) it follows that since, as mentioned earlier,
the two-qutrit isotropic mixed state is entangled if and only if $F > 1/3$ whence Negativity is nonzero, the sum of four PCCs  as given above being greater than $1$
provides necessary and sufficient certification of entanglement. 
Next, we argue that the sum of PCCs given by Eq. (\ref{d3isoNS})  also provides quantification of certified entanglement of the two-qutrit isotropic states
in the following sense.

Now, note that using Eq. (\ref{nqudiso}), one can write Negativity of the two-qutrit isotropic mixed state  for $F > 1/3$
\begin{align}\label{nqut1}
\mathcal{N}(\rho_I(F))= \dfrac{3F -1}{2}.
\end{align}
From the above Eq. (\ref{nqut1}), using Eq. (\ref{d3isoNS}) 
it follows that for $F > 1/3$
\begin{align}\label{SqutI}
\sum^4_{i=1}|\mathcal{C}_{A_iB_i}|=1+3\mathcal{N}(\rho_I(F))
\end{align}
Thus the sum of PCCs is a linear function of Negativity and hence quantifies entanglement in this case.

\subsection{Coloured-noise  mixed with maximally entangled state}

Here we consider two families  of  two-qutrit mixed states having maximally entangled state mixed with two types of coloured-noise. In one of them (labeled A), coloured-noise state has perfect correlation in the computational basis  and in the other type (labeled B), coloured-noise state has 
perfect anti-correlation in the computational basis.

\paragraph*{\bf{Coloured-noise  mixed  states-$A$:}}
Let us write  the  general expression for the  coloured-noise two-qudit maximally entangled state which is a mixture of the two-qudit maximally entangled state  $\ket{\phi_d^+}$ and 
the coloured-noise two-qudit state $1/d\sum^{d-1}_{i=0} \ketbra{ii}{ii}$ given by
\begin{equation}\label{cndme}
\rho_{cc}(p)=p \ketbra{\phi_d^+}{\phi_d^+} +\frac{(1-p)}{d} \sum^{d-1}_{i=0} \ketbra{ii}{ii},
\end{equation}
where $p$ is the mixed parameter, $0 \le p \le 1$.  In Ref. \cite{HMK+16}, experimental verification of entanglement of the above class of states was demonstrated by using the approach based on the sum of mutual information. 
It can be checked that the above class of states is entangled for $p\ne 0$ by using the positive partial transpose criterion \cite{Per96}.
For this class of states, Negativity  as defined in Ref. \cite{VW02}
can be calculated from the partial transposed density matrix is  given by 
\begin{align} \label{Negcnd}
\mathcal{N}(\rho_{cc}(p))=(d-1)\frac{p}{2}.
\end{align}
Since the one-parameter family of 
states given by Eq. (\ref{cndme}) is separable for $p=0$ and for $p\ne0$, $\mathcal{N}(\rho_{cc}(p))>0$, this class of states is entangled if and only if $p>0$.

Let us now consider  the  coloured-noise two-qutrit maximally entangled state, i.e., $\rho_{cc}(p)$ given by Eq. (\ref{cndme}) with $d=3$. 
Let the basis $\{\ket{a_j}\}$ of the  pair of observables $A_1B_1$ in Eq. (\ref{s2pcc})
be the computational basis and the basis $\{\ket{b_j}\}$ of the  pair of observables $A_2B_2$ in Eq. (\ref{s2pcc}) be the generalized
$\sigma_y$ basis.
For this choice of two MUBs, the sum of two PCCs for the  coloured-noise two-qutrit maximally entangled state  is given by
 \begin{equation}
 |\mathcal{C}_{A_1B_1}|+|\mathcal{C}_{A_2B_2}|
 =1+p>1 \quad  \text{iff} \quad  p>0, \label{PCCcc}
\end{equation}
which implies that the above sum of two PCCs being greater than $1$ provides necessary and sufficient criterion for certification of entanglement
of the coloured-noise mixed with two-qutrit maximally entangled state since, as mentioned earlier,  
this class of mixed states is entangled if and only if $p \ne 0$.
See Appendix \ref{AIV} for the derivation of the above expression for the sum of
two PCCs.

It is then readily seen from the expression of Negativity for the coloured-noise two-qutrit maximally entangled state
given by Eq. (\ref{Negcnd}) with $d=3$ that the sum
of PCCs given by Eq. (\ref{PCCcc})  is related to  Negativity as follows:
\begin{equation}
|\mathcal{C}_{A_1B_1}|+|\mathcal{C}_{A_2B_2}|
=1+\mathcal{N}(\rho_{cc}(p))
\end{equation}
thereby providing quantification of entanglement in this case. 
On the other hand, it can be checked  that for any  two noncommuting bases which are \textit{not MUBs} chosen from the set given by Eq. (\ref{csmub3}), the sum of two PCCs being greater than $1$  provides \textit{only} sufficient certification 
of entanglement of the  coloured-noise two-qutrit maximally entangled state.

\paragraph*{\bf{Coloured-noise  mixed  states-$B$:}}
In addition to the above type of mixed state involving coloured noise, we now consider the following
type of state which was first introduced by Eltschka et al in Ref. \cite{ETS15} and later used by Sentis et al in Ref. \cite{SEG+16}.

Let us write as follows the  general expression for this type of mixed state which is a mixture of the two-qudit maximally entangled state  $\ket{\phi_d^+}$ and 
the coloured-noise two-qudit state of the type given by $1/(d(d-1))\sum^{d-1}_{i\ne j=0} \ketbra{ij}{ij}$: 
\begin{equation}\label{cndme1}
\rho_{ac}(p)=p \ketbra{\phi_d^+}{\phi_d^+} +\frac{(1-p)}{d(d-1)} \sum^{d-1}_{i \ne j=0} \ketbra{ij}{ij},
\end{equation}
where $0 \le p \le 1$. 
For this class of states, Negativity  as defined in Ref. \cite{VW02}
can be calculated from the partial transposed density matrix,  given by 
\begin{align} \label{Negcnd1}
\mathcal{N}(\rho_{ac}(p))=\max \Bigg\{ \dfrac{dp -1}{2} , 0 \Bigg \}.
\end{align}

Let us now consider  the  coloured-noise two-qutrit maximally entangled state, i.e., $\rho_{ac}(p)$ given by Eq. (\ref{cndme1}) with $d=3$.
It can be checked that for the two MUBs which are the computational bases and the generalized
$\sigma_y$ basis, the sum of two PCCs for the coloured-noise mixed states given by Eq. (\ref{cndme1}) with $d=3$ is greater than $1$
only when the Negativity is greater than certain value. Therefore, we proceed to check whether the sum of four 
PCCs for this family of mixed states is greater than $1$ for some suitable set of four noncommuting bases 
if and only if the Negativity of the state is nonzero.
We now use the set of four noncommuting  bases (which are not MUBs) given in Eq. (\ref{csmub3}) which we have used 
for certifying and quantifying entanglement of the above mentioned  coloured-noise two-qutrit maximally entangled state using
the sum of $4$ PCCs. For these noncommuting bases,
the sum of four PCCs for the coloured-noise two-qutrit mixed state given by Eq. (\ref{cndme1}) with $d=3$ is given by
 \begin{align}\label{PCCcc1}
 \sum^4_{i=1}|\mathcal{C}_{A_iB_i}|
 &=\frac{9p-1}{2}>1 \quad \text{iff} \quad  p>1/3,
\end{align}
which implies that the above sum of two PCCs is greater than $1$  if and only if the Negativity $\mathcal{N}(\rho_{ac}(p)) \ne 0$.
See Appendix \ref{ACNB} for the derivation of the above expression for the sum of
four PCCs.
It is then readily seen from the expression of Negativity for the coloured-noise two-qutrit maximally entangled state
given by Eq. (\ref{Negcnd1}) with $d=3$ that the sum
of PCCs given by Eq. (\ref{PCCcc1})  is related to  Negativity as follows:
 \begin{align}
 \sum^4_{i=1}|\mathcal{C}_{A_iB_i}|=1+3\mathcal{N}(\rho_{ac}(p)).
\end{align}
thereby providing quantification of certified entanglement, similar to the quantification of entanglement of the two-qutrit isotropic states given by Eq. (\ref{SqutI}).

\subsection{Werner states}
In Ref. \cite{Wer89}, Werner introduced a class of mixed two-qudit states for which there are separable as well as entangled subsets, the latter containing states for which 
local realist model exists.  
These mixed two-qudit states are called Werner states. Here we consider a particular form 
of such a state in any dimension which is a convex mixture of the projector onto the antisymmetric space
and white noise \cite{QVC+15} given by
\be\label{wernerd}
\rho_W(p)=\frac{p}{d(d-1)}2 P_{anti} +\frac{(1-p)}{d^2} \mathbb{I},
\ee
where 
\be
 1-\frac{2d}{d+1} \le p \le 1, \nonumber
\ee
and
\be
 P_{anti}=\frac{1}{2}\left(\mathbb{I}- \sum^{d-1}_{ij=0}\ketbra{i}{j} \otimes \ketbra{j}{i}\right) \nonumber
\ee
which is the projector onto the anti-symmetric space. Note that for $d=2$, the above class of states is a mixture of the maximally entangled state and white noise.

For the two-qudit Werner state $\rho_W(p)$ given by Eq. (\ref{wernerd}), Negativity as defined in Ref. \cite{VW02} can be computed from the partial transposed density matrix
and is given by 
\begin{align}\label{nqudwer}
\mathcal{N}(\rho_W(p))=\max \Bigg\{ \dfrac{(d+1)p -1}{d^2} , 0 \Bigg \}		
\end{align}
which is nonzero if and only if $p>1/(d+1)$. Also, note that the two-qudit Werner state $\rho_W(p)$ given by Eq. (\ref{wernerd}) is entangled
if and only if $p>1/(d+1)$ \cite{Wer89,QVC+15}. Therefore, it follows that the Negativity of this class of states as given by Eq. (\ref{nqudwer}) provides the necessary and sufficient
quantification of entanglement for any $d$.
We may note here that for $d \ge 3$, the existence of an entanglement witness for such class of states
which is experimentally measurable has been shown \cite{SLD15} but the quantification of certified entanglement of the Werner states  has remained uninvestigated. Thus, in this context, the following procedure of entanglement characterization using the measurable PCCs is of particular significance.

Let us now consider  the two-qutrit Werner state, i.e., $\rho_W(p)$ given by Eq. (\ref{wernerd}) with $d=3$. 
 For the four noncommuting  bases (which are not MUBs) given in Eq. (\ref{csmub3}), i.e.,
$A_1=B_1=\sum_j a_j \ketbra{a_j}{a_j}$, $A_2=B_2=\sum_j b_j \ketbra{b_j}{b_j}$, 
$A_3=B_3=\sum_j e_j \ketbra{e_j}{e_j}$ and $A_4=B_4=\sum_j g_j \ketbra{g_j}{g_j}$,
the sum of four PCCs for the two-qutrit Werner state is given by
 \begin{align}\label{sPw}
 \sum^4_{i=1}|\mathcal{C}_{A_iB_i}|
 &=2|p|>1 \quad \text{iff} \quad  p>1/2.
\end{align}
See Appendix \ref{AV} for the derivation of the above expression.
Since, as mentioned earlier, the Werner states given by Eq. (\ref{wernerd}) with $d=3$ are entangled for $p>1/4$,
it follows from Eq. (\ref{sPw}) that the  sum of four PCCs  being greater than $1$ provides sufficient criterion for the certification of entanglement
of the state given by Eq. (\ref{nqudwer}) with $d=3$. Interestingly, it is found that the expression for the sum of four PCCs obtained in Eq. (\ref{sPw}) for the two-qutrit Werner states can also be obtained by the set of four MUBs given by Eq. (\ref{csmubd=4}) in Appendix \ref{MUBsd35}.
Next, we argue that the sum of PCCs given by Eq. (\ref{sPw})  also provides quantification of certified entanglement of the Werner states.

\begin{widetext}
 \begin{center}
\begin{figure}[!t]
\includegraphics[scale=0.9]{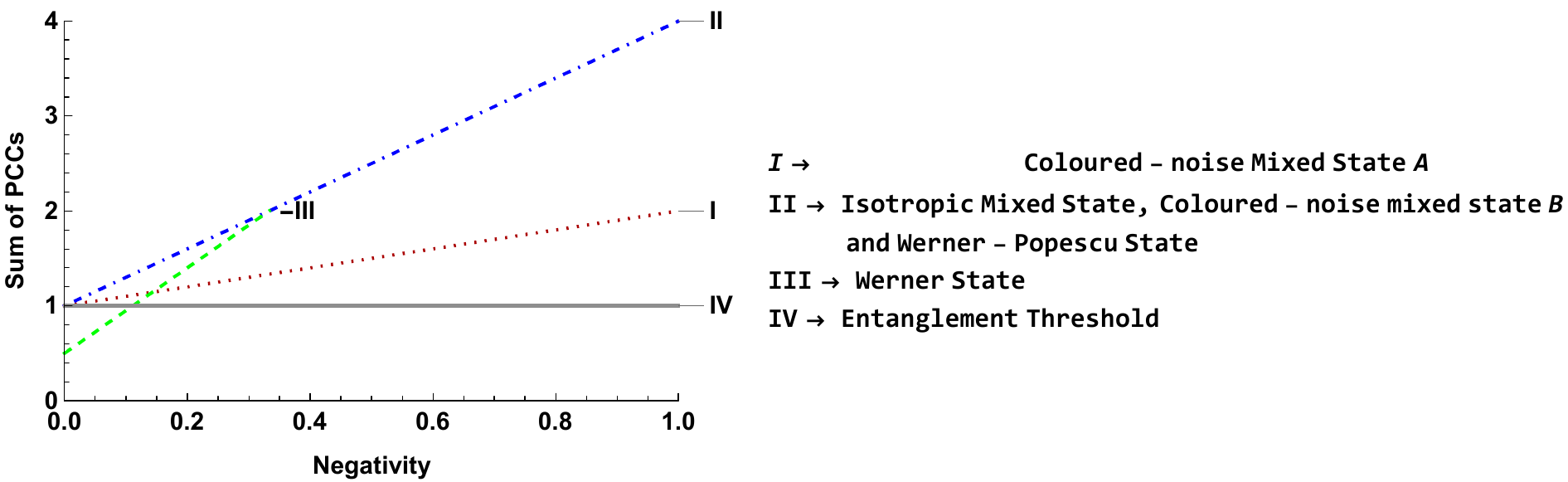}
\caption{\footnotesize For $d=3$, the sum of PCCs is plotted as a function of negativity  for the  six  families of two-qudit states indicated in the right hand side. 
 The dotted line (I) corresponds to the sum of two PCCs versus negativity for the  coloured-noise mixed state A given by Eq. (\ref{PCCcc})  in the text.
 The dot-dashed line (II) denotes the sum of four PCCs versus negativity for the isotropic mixed state,  
 coloured-noise mixed state B and Werner-Popescu state given by Eqs. (\ref{d3isoNS}), (\ref{PCCcc1}) and (\ref{d3WPNS}) respectively. 
 The dashed line (III) indicates the sum of four PCCs versus negativity for the Werner states given by Eq. (\ref{SqutW}).
The horizontal line (IV)  specifies entanglement threshold above which the states are entangled.}
\label{PlotD=3}
\end{figure}
\end{center}
\end{widetext}

Note that using  Eq. (\ref{nqudwer}), Negativity of the two-qutrit Werner state for $p > 1/4$ given by
\begin{align}\label{nWqut1}
\mathcal{N}(\rho_W(p))= \dfrac{4p -1}{9}.	
\end{align}
From the above Eq. (\ref{nWqut1}), using Eq. (\ref{sPw}) 
it follows that for $p > 1/4$
\begin{align}\label{SqutW}
\sum^4_{i=1}|\mathcal{C}_{A_iB_i}|=\frac{1+9\mathcal{N}(\rho_W(p))}{2}.	
\end{align}
Thus the sum of PCCs is a linear function of Negativity and hence quantifies entanglement in this case.

\subsection{Werner-Popescu states}

The so called Werner-Popescu state \cite{HH99,Pop94} in arbitrary dimension $d$ which is a convex mixture of the maximally entangled pure two-qudit state
and white noise is given by
\begin{equation}{\label{WPd}}
	\rho_{WP}(p) 
	= \dfrac{1-p}{d^2}  \mathbb{I} + \, p \, \ketbra{\phi_d^+}{\phi_d^+},
	\end{equation}
which has also been discussed elsewhere, for instance, in Ref. \cite{HMK+16}.
For $d=2$, Werner-Popescu states become same as the Werner states up to local unitary.

Note that the isotropic mixed state given by Eq. (\ref{Isomixedsattedef}) can be written in the form of $\rho_{WP}(p)$ given above with $F=\frac{(d^2-1)p+1}{d^2}$, 
for $F \ge 1/d^2$ since $p$ lies between $0$ and $1$.
Now, $F>1/d$ implies $p>1/(d+1)$ and, as mentioned earlier, the two-qudit isotropic state is entangled if and only if $F>1/d$. It thus follows that the two-qudit Werner-Popescu state $\rho_{WP}(p)$ given by Eq. (\ref{WPd}) is entangled
if and only if $p>1/(d+1)$ \cite{HH99}.

Let us now consider the two-qutrit Werner-Popescu state, i.e.,  $\rho_{WP}(p)$ given by Eq. (\ref{WPd}) with $d=3$.
For the choice of four noncommuting bases (not MUBs) given by Eq. (\ref{csmub3}), the sum of four PCCs for the two-qutrit Werner-Popescu state is given by
 \begin{align}\label{d3WPNS}
 \sum^4_{i=1}|\mathcal{C}_{A_iB_i}|
 &=4p>1 \quad \text{iff} \quad  p>1/4. 
 \end{align}
See Appendix \ref{AWP} for the derivation of the above expression for the sum of
four PCCs. Since, as mentioned earlier,
the Werner-Popescu state given by Eq. (\ref{WPd}) with $d=3$ is entangled if and only if $p > 1/4$, 
the sum of four PCCs  as given above being greater than $1$
provides necessary and sufficient certification of entanglement.

 While  the above demonstration of necessary and sufficient certification of entanglement has been in terms of four noncommuting bases which are \textit{not MUBs},  it can be checked that for the set of four MUBs which include the 
computational basis and generalized $\sigma_x$-basis, the sum of four PCCs being greater than $1$  provides \textit{only} sufficient certification 
of entanglement of the two-qutrit Werner-Popescu states.
Next, we argue that the sum of PCCs given by Eq. (\ref{d3WPNS})  also provides quantification of certified entanglement of the two-qutrit Werner-Popescu  states
in the following sense.

For the  two-qutrit Werner-Popescu  state $\rho_{WP}(p)$ given by Eq. (\ref{WPd}) with $d=3$, Negativity as defined in Ref. \cite{VW02} can be computed from the partial transposed density matrix
and is given by 
\begin{align}\label{nWP3}
\mathcal{N}(\rho_{WP}(p))=\max \Bigg\{ \dfrac{4p -1}{3} , 0 \Bigg \}	
\end{align}
which is nonzero if and only if $p>1/4$. Interestingly, the two-qutrit isotropic mixed state $\rho_{WP}(p)$ is entangled
if and only if $p>1/4$ \cite{HH99}. Therefore, it follows that the Negativity of this class of states as given by Eq. (\ref{nWP3}) provides the necessary and sufficient
quantification of entanglement.

Now, note that using Eq. (\ref{nWP3}), one can write Negativity of the two-qutrit Werner-Popescu  state $\rho_{WP}(p)$  for $p > 1/4$
\begin{align}\label{nWP31}
\mathcal{N}(\rho_{WP}(p))= \dfrac{4p -1}{3}.
\end{align}
From the above Eq. (\ref{nWP31}), using Eq. (\ref{d3WPNS}) 
it follows that for $p > 1/4$
\begin{align}\label{SqutWP}
\sum^4_{i=1}|\mathcal{C}_{A_iB_i}|=1+3\mathcal{N}(\rho_{WP}(p))
\end{align}
Thus the sum of PCCs is a linear function of Negativity and hence quantifies entanglement in this case.

Next, we proceed to investigate to what extent the approach using PCCs can provide certification and quantification of entanglement for the pure states and the above classes of states for $d=4$ and $5$ as well as pure states.

\section{Two-qudit states for $d=4$ and $d=5$} 
\subsection{Pure states}
\paragraph*{\bf{For d=4:}}
Let us consider the pure two-qudit state of dimension $d=4$ of the form
\begin{equation}
 \ket{\psi_4}=c_0\ket{00}+c_1\ket{11}+c_2\ket{22}+c_3\ket{33} \label{q4Sc}
\end{equation}
where $0 \le c_0,c_1,c_2, c_3 \le 1$ and $\sum^3_{i=0} c^2_i=1$. 
For the above class of states, the  expression for Negativity is given by 
\begin{equation}\label{pd4neg}
\mathcal{N}(\ket{\psi_4})=c_0c_1+c_0c_2+c_0c_3+c_1c_2+c_1c_3+c_2c_3
\end{equation}

The above expression can be obtained from the general formula for Negativity for a pure two-qudit state $|\psi_{d}\rangle $ given by \citep{ETS15}

\begin{equation} \label{neweq}
\mathcal{N}(|\psi_{d}\rangle)=\sum^{d-1}_{p\neq q=0, p\rangle q} C_{p}C_{q}
\end{equation}

where $\ket\psi_{d}$ is of the Schmidt decomposition form

\begin{equation}\label{neweq1}
\ket{\psi_{d}} = \sum^{d-1}_{i=0} C_{i} \ket{ii}
\end{equation}

In Sec. II A, the generalized $\sigma_z$ basis and the generalized $\sigma_y$ basis have been defined for any dimension $d\ge 3$.
For this choice of two MUBs in the case  $d=4$,
the sum of two PCCs for the pure two-qudit states of dimension $d=4$ given by Eq. (\ref{q4Sc})
can be shown to be given by
\begin{align}
 &|\mathcal{C}_{A_1B_1}|+|\mathcal{C}_{A_2B_2}| \nonumber \\
 &=1+\frac{9 c_2 c_3 + c_1 (9 c_2 + 2 c_3) + 
 c_0 (9 c_1 + 2 c_2 + 9 c_3)}{10} \label{sump4}
\end{align}
where $A_1=B_1=\sum_j a_j \ketbra{a_j}{a_j}$ and $A_2=B_2=\sum_j b_j \ketbra{b_j}{b_j}$, with $\{\ket{a_j}\}$
and $\{\ket{b_j}\}$ being the generalized $\sigma_z$ basis and the generalized $\sigma_y$ basis, respectively,
and the eigenvalues are given by $a_0=b_0=+2$, $a_1=b_1=+1$, $a_2=b_2=-1$ and $a_3=b_3=-2$.
From Eqs. (\ref{pd4neg}) and (\ref{sump4}) it follows that if and only if any two of $c_i$'s  are nonzero,
then Negativity is nonzero 
as well as the sum of PCCs given by Eq. (\ref{sump4}) is greater than $1$. Now, since a pure two-qudit state is entangled
if and only if Negativity is nonvanishing, we can argue that 
for the pure two-qudit states of dimension $d=4$, the sum of PCCs being greater than $1$ provides necessary
and sufficient certification of entanglement. Note that
the  sum of PCCs given by Eq. (\ref{sump4}) attains the algebraic maximum of $2$ for the maximally 
entangled state for which all $c_i$s in Eq. (\ref{sump4}) are equal to $1/\sqrt{4}$.

As regards quantification of entanglement, it can be checked that the sum
of PCCs given by Eq. (\ref{sump4}) is related to Negativity as follows:
\begin{align}
|\mathcal{C}_{A_0B_0}|+|\mathcal{C}_{A_1B_1}|=1+\frac{9\mathcal{N}(\ket{\psi_4})-7\chi}{10}  \label{sumpN4}
\end{align}
where $\chi=c_0c_2+c_1c_3$ which takes value in the interval $0 \le \chi \le1/2$.
The relationship  between the sum of PCCs and Negativity given above
implies that for any class of pure states for which the quantity 
$\chi$ takes a constant value  $c$, the sum of PCCs given by Eq. (\ref{sump4})
is a monotonic function of Negativity. This means that for any pair of pure states within a class of pure states for which
$\chi=c$, higher value of the sum of PCCs  given by Eq. (\ref{sumpN4}) always implies higher degree of entanglement.

For the more general class of pure states
given by Eq. (\ref{q4Sc}), whether the sum of PCCs for any other possible two MUBs  is   a monotonic
function of Negativity is a critical issue. It has been checked that the optimization of the sum of two PCCs for this class of pure states
would not lead to such a linear relationship with the Negativity which ensures that the sum of PCCs takes the maximum value of $2$ for the 
maximally entangled state. The sought after linear relationship between the sum of PCCs and Negativity should read  as 
$|\mathcal{C}_{A_0B_0}|+|\mathcal{C}_{A_1B_1}|=1+2/3\mathcal{N}$. It has been found that for the nonmaximally entangled pure states, 
the sum of PCCs that has this form take lower value than the sum of PCCs having the form given by Eq. (\ref{sumpN4}). Therefore, 
optimization of the sum of PCCs for the pure states in $d=4$ over all possible two MUBs cannot lead to necessary and sufficient certification
as well as quantification of entanglement of the pure states.

We have also done a thorough numerical study which shows  that for any two MUBs, one of  which is the computational basis, 
the sum of two PCCs for the pure states given by Eq. (\ref{q4Sc})  does not have the relationship 
with Negativity that is required for quantification of certified entanglement (see Appendix \ref{LRNd=4} for the relevant discussion of this numerical study).
\paragraph*{\bf{For d=5:}}
Let us consider the general pure two-qudit state of dimension $d=5$ given by
\begin{equation}
 \ket{\psi_5}=c_0\ket{00}+c_1\ket{11}+c_2\ket{22}+c_3\ket{33}+c_4 \ket{44} \label{q5Sc}
\end{equation}
where $0 \le c_0,c_1,c_2, c_3,c_4 \le 1$ and $\sum^4_{i=0} c^2_i=1$. 
For the above class of states, the general expression for Negativity given by Eq. (\ref{neweq}) reduces to 
\begin{align}\label{pd5neg}
\mathcal{N}(\ket{\psi_5})&=c_0(c_1+c_2+c_3+c_4)+c_1(c_2+c_3+c_4) \nonumber \\
&+c_2(c_3+c_4)+c_3c_4
\end{align}

For the two MUBs which are taken to be the generalized $\sigma_z$ basis and the generalized $\sigma_y$ basis  for  $d=5$,
the sum of two PCCs for the pure two-qudit states of dimension $d=5$ given by Eq. (\ref{q5Sc})
can be shown to be given by
\begin{align}
 &|\mathcal{C}_{A_1B_1}|+|\mathcal{C}_{A_2B_2}| \nonumber \\
 &=1+\frac{5+\sqrt{5}}{10}\left(c_0c_1+c_0c_4+c_1c_2+c_2c_3+c_3c_4\right) \nonumber \\
 &+\frac{5-\sqrt{5}}{10}\left(c_0c_2+c_0c_3+c_1c_3+c_1c_4+c_2c_4\right) \label{sump5}
\end{align}
where, similar to that mentioned for $d=4$, we have taken $A_1=B_1=\sum_j a_j \ketbra{a_j}{a_j}$ and $A_2=B_2=\sum_j b_j \ketbra{b_j}{b_j}$, with $\{\ket{a_j}\}$
and $\{\ket{b_j}\}$ being the generalized $\sigma_z$ basis and the generalized $\sigma_y$ basis respectively 
and the eigenvalues are given by $a_0=b_0=+2$, $a_1=b_1=+1$, $a_2=b_2=0$, $a_3=b_3=-1$ and $a_3=b_3=-2$.
The above sum of PCCs given by Eq. (\ref{sump5}) attains the algebraic maximum of $2$ for the maximally 
entangled state for which all $c_i$s in Eq. (\ref{sump5}) are equal to $1/\sqrt{5}$.

Now, from Eqs. (\ref{pd5neg}) and (\ref{sump5}) it follows that if and only if any two of $c_i$'s  are nonzero,
then Negativity is nonzero 
as well as the sum  PCCs given by Eq. (\ref{sump4}) is greater than $1$. 
Thus, for the pure two-qudit states of dimension $d=5$, the sum of PCCs being greater than $1$ provides necessary
and sufficient certification of entanglement.

As regards quantification of entanglement, it can be checked that the sum
of PCCs given in Eq. (\ref{sump5}) is related to the Negativity as follows:
\begin{align}
&|\mathcal{C}_{A_0B_0}|+|\mathcal{C}_{A_1B_1}|=1+\frac{(5+\sqrt{5})\mathcal{N}(\ket{\psi_5})-2\sqrt{5}\chi}{10}   \label{sumpN5}
\end{align}
where $\chi=c_0c_2+c_0c_3+c_1c_3+c_1c_4+c_2c_4$ which takes value in the interval $0 \le \chi \le 1$. 
Similar to the case of $d=4$ pure states, the relationship  between the sum of PCCs and Negativity given above
implies that for any pair of pure states drawn from a class of pure states for which the quantity
$\chi$ takes a constant value $c$, higher value of the sum of the PCCs given by Eq. (\ref{sumpN5}) always 
implies higher value of entanglement.

For the more general class of pure states
given by Eq. (\ref{q5Sc}), in this case, too, similar to $d=4$, by optimizing over all possible  
two MUBs, one cannot obtain an expression for the sum of two PCCs  which  is linearly related with Negativity.
As in the case of $d=4$ pure states, it has also been  checked by thorough numerical search over all possible MUBs, one of which is 
the computational basis, that the approach based on the sum of
two PCCs for the pure states in $d=5$ does not provide quantification of certified entanglement, as in the
case of pure states in $d=4$.
\subsection{Isotropic mixed states}

Now, following the  procedure of entanglement characterization using the measurable PCCs as shown for two-qutrit isotropic mixed states, 
we now proceed to address the $d=4$ and $d=5$ cases. 
\paragraph*{\bf{For d=4:}}

Now, to certify entanglement of the isotropic mixed state given by Eq. (\ref{Isomixedsattedef}) in dimension $d=4$, 
we use the sum of five PCCs $\sum^5_{i=1}|\mathcal{C}_{A_iB_i}|$, 
where $A_1=B_1=\sum_j a_j \ketbra{a_j}{a_j}$, $A_2=B_2=\sum_j b_j \ketbra{b_j}{b_j}$, 
$A_3=B_3=\sum_j e_j \ketbra{e_j}{e_j}$, $A_4=B_4=\sum_j g_j \ketbra{g_j}{g_j}$ and $A_5=B_5=\sum_j k_j \ketbra{k_j}{k_j}$
with the eigenvalues $a_0=b_0=e_0=g_0=k_0=+2$, $a_1=b_1=e_1=g_1=k_1=+1$, $a_2=b_2=e_2=g_2=k_2=-1$ and $a_3=b_3=e_3=g_3=k_3=-2$.
Detailed expressions for the five bases corresponding to these observables are given by Eq. (\ref{csmub4}) in Appendix \ref{NCB45}.
For this choice of five mutually unbiased bases, the sum of five PCCs is given by
 \begin{align}\label{d4isoNS}
 \sum^5_{i=1}|\mathcal{C}_{A_iB_i}|
 &=\frac{|16F-1|}{3}>1 \quad \text{iff} \quad  F>1/4. 
 \end{align}
Since the isotropic mixed state (given by Eq. (\ref{Isomixedsattedef}))  is entangled for $F>1/4$ for dimension $d=4$,
it follows that the  sum of five PCCs given by Eq. (\ref{d4isoNS}) being greater than $1$ provides necessary and sufficient criterion for the certification of entanglement
of the isotropic mixed state given by Eq. (\ref{Isomixedsattedef}) in dimension $d=4$.
Next, we argue that the sum of PCCs given by Eq. (\ref{d4isoNS})  also provides quantification of certified entanglement of the isotropic mixed state given by Eq.   (\ref{Isomixedsattedef}) in dimension $d=4$.

Note that using Eq. (\ref{nqudiso}), one can write Negativity of the entangled isotropic mixed state (given by Eq. (\ref{Isomixedsattedef})) in dimension $d=4$
for $F > 1/4$ given by
\begin{align}\label{nqud4}
\mathcal{N}(\rho_I(p))= \dfrac{4F -1}{2}.
\end{align}
From the above Eq. (\ref{nqud4}), using Eq. (\ref{d4isoNS}) 
it follows that for $F > 1/4$
\begin{align}\label{Squd4I}
\sum^5_{i=1}|\mathcal{C}_{A_iB_i}|=1+\frac{8}{3}\mathcal{N}(\rho_I(p))
\end{align}
Thus the sum of PCCs is a linear function of Negativity and hence quantifies entanglement in this case.

\paragraph*{\bf{For d=5:}}
Similarly, 
now, to certify entanglement of the isotropic mixed state given by Eq. (\ref{Isomixedsattedef}) in dimension $d=5$, 
we use the sum of six PCCs $\sum^6_{i=1}|\mathcal{C}_{A_iB_i}|$, 
where $A_1=B_1=\sum_j a_j \ketbra{a_j}{a_j}$, $A_2=B_2=\sum_j b_j \ketbra{b_j}{b_j}$, 
$A_3=B_3=\sum_j e_j \ketbra{e_j}{e_j}$, $A_4=B_4=\sum_j g_j \ketbra{g_j}{g_j}$, $A_5=B_5=\sum_j k_j \ketbra{k_j}{k_j}$
and $A_6=B_6=\sum_j l_j \ketbra{l_j}{l_j}$  with the eigenvalues $a_0=b_0=e_0=g_0=k_0=l_0=+2$, $a_1=b_1=e_1=g_1=k_1=l_1=+1$, 
$a_2=b_2=e_2=g_2=k_2=l_2=0$, $a_2=b_2=e_2=g_2=k_2=l_2=-1$ and $a_3=b_3=e_3=g_3=k_3=l_3=-2$.
Detailed expressions for the six bases corresponding to these observables are given by Eq. (\ref{ncb5}) in Appendix \ref{NCB45}.
For this choice of six noncommuting bases which are not MUBs, the sum of six PCCs is given by
 \begin{align}\label{d5isoNS}
 \sum^6_{i=1}|\mathcal{C}_{A_iB_i}|
 &=\frac{|25F-1|}{4}>1 \quad \text{iff} \quad  F>1/5 
 \end{align}
Since the isotropic mixed state given by Eq. (\ref{Isomixedsattedef})  is entangled for $F>1/5$ for dimension $d=4$,
it follows that the  sum of six PCCs given by Eq. (\ref{d5isoNS}) being greater than $1$ provides necessary and sufficient criterion for the certification of entanglement
of the isotropic mixed state given by Eq. (\ref{Isomixedsattedef}) in dimension $d=5$.
Similar to the case $d=4$, we now argue that the sum of PCCs given by Eq. (\ref{d5isoNS})  also provides quantification of certified entanglement of the isotropic mixed state given by Eq. (\ref{Isomixedsattedef}) in dimension $d=5$
in the following sense.

Now, note that using Eq. (\ref{nqudiso}), one can write Negativity of the entangled isotropic mixed state (given by Eq. (\ref{Isomixedsattedef})) in dimension $d=5$
for $p > 1/5$ given by
\begin{align}\label{nqud5}
\mathcal{N}(\rho_I(F))= \dfrac{5F -1}{2}.
\end{align}
From  Eq. (\ref{nqud5}), using Eq. (\ref{d5isoNS}) 
it follows that for $F > 1/5$
\begin{align}\label{Squd5I}
\sum^5_{i=1}|\mathcal{C}_{A_iB_i}|=1+\frac{5}{2}\mathcal{N}(\rho_I(F))
\end{align}
Thus the sum of PCCs is a linear function of Negativity and hence quantifies entanglement in this case.

\subsection{Coloured-noise  mixed with maximally entangled state}
 Here we consider two types of  coloured-noise  state mixed with the maximally entangled two-qudit state given by Eqs. (\ref{cndme}) and (\ref{cndme1}) which are abbreviately
called coloured-noise  mixed  states-$A$ and coloured-noise  mixed  states-$B$ respectively.
\paragraph*{\bf{Coloured-noise  mixed  states-$A$:}}

\paragraph*{\bf{For d=4:}}
In order to certify entanglement of the coloured-noise two-qudit maximally entangled state (given by Eq. (\ref{cndme})) in dimension $d=4$ as in the case for $d=3$, 
we use the criterion given by Eq. (\ref{s2pcc}). 
Let the basis $\{\ket{a_j}\}$ of the  pair of observables $A_1B_1$ in Eq. (\ref{s2pcc})
be the computational basis and the basis $\{\ket{b_j}\}$ of the  pair of observables $A_2B_2$ in Eq. (\ref{s2pcc}) be the generalized
$\sigma_y$ basis.
For this choice of two MUBs, the sum of two PCCs computed for the state given by Eq. (\ref{cndme}) for $d=4$ is given by
 \begin{equation}
 |\mathcal{C}_{A_1B_1}|+|\mathcal{C}_{A_2B_2}|
 =1+p>1 \quad  \text{iff} \quad  p>0, \label{PCCcc4}
\end{equation}
from which it follows  that the above sum of two PCCs being greater than $1$ provides necessary and sufficient criterion for certification of entanglement
of the coloured-noise mixed with two-qudit maximally entangled state in dimension $d=4$ since, as mentioned earlier,  
this class of mixed states is entangled if and only if $p \ne 0$.
It is also readily seen from Eqs. (\ref{PCCcc4}) and (\ref{Negcnd}) for $d=4$ that the sum
of PCCs  is related to  Negativity as follows:
\begin{equation} \label{PCCcc4N}
|\mathcal{C}_{A_1B_1}|+|\mathcal{C}_{A_2B_2}|
=1+\frac{2}{3}\mathcal{N}(\rho_{cc}(p))
\end{equation}
thereby providing quantification of entanglement in this case. 

\paragraph*{\bf{For d=5:}}
Similar to the above case, we consider  the basis $\{\ket{a_j}\}$ of the  pair of observables $A_1B_1$ in Eq. (\ref{s2pcc}) to
be the computational basis and the basis $\{\ket{b_j}\}$ of the  pair of observables $A_2B_2$ in Eq. (\ref{s2pcc}) to be the generalized
$\sigma_y$ basis.
For this choice of two MUBs, the sum of two PCCs computed using the state given by Eq. (\ref{cndme}) for $d=5$ is given by
 \begin{equation}
 |\mathcal{C}_{A_1B_1}|+|\mathcal{C}_{A_2B_2}|
 =1+p>1 \quad  \text{iff} \quad  p>0, \label{PCCcc5}
\end{equation}
which shows, similar to the earlier case for $d=4$, that the above sum of two PCCs being greater than $1$ provides necessary and sufficient criterion for certification of entanglement
of the coloured-noise mixed with two-qudit maximally entangled state in dimension $d=5$. 
It is then also seen from Eqs. (\ref{PCCcc5}) and (\ref{Negcnd}) for $d=5$ that the sum
of two PCCs  is related to  Negativity as follows:
\begin{equation} \label{PCCccN5}
|\mathcal{C}_{A_1B_1}|+|\mathcal{C}_{A_2B_2}|
=1+\frac{1}{2}\mathcal{N}(\rho_{cc}(p))>1 \quad  \text{iff} \quad  \mathcal{N}>0,
\end{equation}
thereby providing quantification of entanglement in this case.

\paragraph*{\bf{Coloured-noise  mixed  states-$B$:}}

\paragraph*{\bf{For d=4:}}
Now, to certify entanglement of the coloured-noise mixed state given by Eq. (\ref{cndme1}) in dimension $d=4$, 
we use the sum of five PCCs $\sum^5_{i=1}|\mathcal{C}_{A_iB_i}|$
for the five noncommuting bases given by Eq. (\ref{csmub4}) in Appendix \ref{NCB45} which we have used in the case of entanglement certification of isotropic
mixed states in $d=4$.
This sum of   PCCs takes the following expression for  the coloured-noise mixed state given by Eq. (\ref{cndme1}) in dimension $d=4$:
 \begin{align}\label{d4cnpcc}
 \sum^5_{i=1}|\mathcal{C}_{A_iB_i}|
 &=\frac{|16p-1|}{3}>1 \quad \text{iff} \quad  p>1/4. 
 \end{align}
The coloured-noise mixed state (given by Eq. (\ref{cndme1})) has Negativity for dimension $d=4$ given by
\begin{align}\label{nqudcn4}
\mathcal{N}(\rho_{ac}(p))= \dfrac{4p -1}{2},
\end{align}
for $p \ge 1/4$ which implies  that the  sum of five PCCs given by Eq. (\ref{d4cnpcc}) is greater than $1$ if and only if the Negativity of the state is nonzero.
Next, we argue that the sum of PCCs given by Eq. (\ref{nqudcn4})  also provides quantification of certified entanglement of the  mixed state given by Eq. (\ref{cndme1}) in dimension $d=4$.
From the above Eq. (\ref{nqudcn4}), using Eq. (\ref{d4cnpcc}) 
it follows that for $p > 1/4$
\begin{align}\label{d4cnpccN}
\sum^5_{i=1}|\mathcal{C}_{A_iB_i}|=1+\frac{8}{3}\mathcal{N}(\rho_{ac}(p))
\end{align}
Thus the sum of PCCs is a linear function of Negativity and hence quantifies certified entanglement.

\paragraph*{\bf{For d=5:}}
Similarly, 
now, to certify entanglement of  the coloured-noise mixed state given by Eq. (\ref{cndme1}) in dimension $d=5$, 
we use the sum of six PCCs $\sum^6_{i=1}|\mathcal{C}_{A_iB_i}|$ 
for the six 
noncommuting bases given by Eq. (\ref{ncb5}) in Appendix \ref{NCB45} with the eigenvalues $a_0=b_0=e_0=g_0=k_0=l_0=+2$, $a_1=b_1=e_1=g_1=k_1=l_1=+1$, 
$a_2=b_2=e_2=g_2=k_2=l_2=0$, $a_3=b_3=e_3=g_3=k_3=l_3=-1$ and $a_4=b_4=e_4=g_4=k_4=l_4=-2$.
This sum of six PCCs takes the following expression for the coloured-noise mixed state given by Eq. (\ref{cndme1}) in $d=5$:
 \begin{align}\label{d5cnpcc}
 \sum^6_{i=1}|\mathcal{C}_{A_iB_i}|
 &=\frac{|25p-1|}{4}>1 \quad \text{iff} \quad  p>1/5 
 \end{align}
The coloured-noise mixed state (given by Eq. (\ref{cndme1})) has Negativity for dimension $d=5$ given by
\begin{align}\label{nqudcn5}
\mathcal{N}(\rho_{ac}(p))= \dfrac{5p -1}{2}.
\end{align}
for $p \ge 1/5$, which implies  that the  sum of six PCCs given by Eq. (\ref{d5cnpcc}) is greater than $1$ if and only if the Negativity of the state is nonzero.
Next, we argue that the sum of PCCs given by Eq. (\ref{nqudcn5})  also provides quantification of certified entanglement of the  mixed state given by Eq. (\ref{cndme1}) in dimension $d=5$.
From the above Eq. (\ref{nqudcn5}), using Eq. (\ref{d5cnpcc}) 
it follows that for $p > 1/5$
\begin{align}\label{d5cnpccN}
\sum^6_{i=1}|\mathcal{C}_{A_iB_i}|=1+\frac{5}{2}\mathcal{N}(\rho_{ac}(p))
\end{align}
Thus the sum of PCCs is a linear function of Negativity and hence quantifies certified entanglement.

\begin{widetext}
 \begin{center}
\begin{figure}[!t]
\includegraphics[scale=0.9]{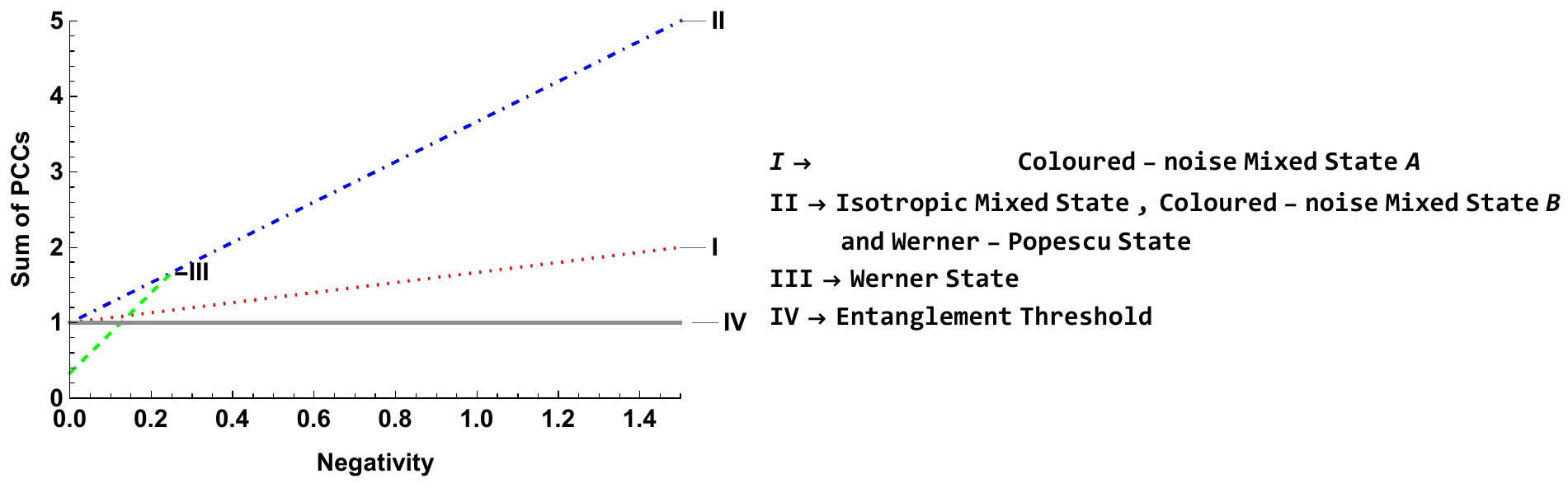}
\caption{\footnotesize For $d=4$, the sum of PCCs is plotted as a function of negativity  for the  six  families of two-qudit states   indicated in the right hand side. 
 The dotted line (I) corresponds to the sum of two PCCs versus negativity for the coloured-noise mixed state A given by Eq. (\ref{PCCcc4N}) in the text.
The dot-dashed line (II) denotes the sum of five PCCs versus negativity for the isotropic mixed state, coloured-noise mixed state B
and Werner-Popescu state given by Eqs. (\ref{Squd4I}), (\ref{d4cnpccN}) and (\ref{d4WPNS}) respectively.
The dashed line (III) indicates the sum of five PCCs versus negativity for the Werner states given by Eq. (\ref{SqudW4}).
The horizontal line (IV) specifies entanglement threshold above which the states are entangled.}
\label{PlotD=4}
\end{figure}
\end{center}
\end{widetext}

\subsection{Werner states}

Now, following the  procedure of entanglement characterization using the PCCs as shown for two-qutrit Werner states, we now proceed to address the $d=4$ and $d=5$ cases. 

\paragraph*{\bf{For d=4:}}
In order to certify entanglement of the Werner state given by Eq. (\ref{wernerd}) in dimension $d=4$, 
we invoke the sum of five PCCs $\sum^5_{i=1}|\mathcal{C}_{A_iB_i}|$, 
where $A_1=B_1=\sum_j a_j \ketbra{a_j}{a_j}$, $A_2=B_2=\sum_j b_j \ketbra{b_j}{b_j}$, 
$A_3=B_3=\sum_j e_j \ketbra{e_j}{e_j}$, $A_4=B_4=\sum_j g_j \ketbra{g_j}{g_j}$ and $A_5=B_5=\sum_j k_j \ketbra{k_j}{k_j}$.
Using the five noncommuting  bases which are  MUBs given by Eq. (\ref{csmub4}) in Appendix \ref{NCB45}
with the eigenvalues $a_0=b_0=e_0=g_0=k_0=+2$, $a_1=b_1=e_1=g_1=k_1=+1$, $a_2=b_2=e_2=g_2=k_2=-1$ and $a_3=b_3=e_3=g_3=k_3=-2$,
the sum of five PCCs in this case computed for the state given by Eq. (\ref{wernerd}) for $d=4$ is as follows
 \begin{align}\label{sPw4}
 \sum^5_{i=1}|\mathcal{C}_{A_iB_i}|
 &=\frac{5}{3}|p|>1 \quad \text{iff} \quad  p>3/5.
\end{align}
Since the Werner states given by Eq. (\ref{wernerd}) are entangled for $p>1/5$ in dimension $d=4$,
it follows that the  sum of five PCCs given by Eq. (\ref{sPw4}) being greater than $1$ provides sufficient criterion for the certification of entanglement
of the Werner states  in dimension $d=4$.
Next, we argue that the sum of PCCs given by Eq. (\ref{sPw4})  also provides quantification of certified entanglement of the Werner states
in the following sense.

For the two-qudit Werner state $\rho_W(p)$ given by Eq. (\ref{wernerd}) in dimension $d=4$, Negativity as defined in Ref. \cite{VW02}  computed from the partial transposed density matrix is given by 
\begin{align}\label{nWqud4}
\mathcal{N}(\rho_W(p))=\max \Bigg\{ \dfrac{5p -1}{16} , 0 \Bigg \}	
\end{align}
which is nonzero if and only if $p>1/5$.  Now, note that using Eq. (\ref{nWqud4}), one can write for $p > 1/5$
\begin{align}\label{nWqud4'}
\mathcal{N}(\rho_W(p))= \dfrac{5p -1}{16}.
\end{align}
From the above Eq. (\ref{nWqud4'}), using Eq. (\ref{sPw4}) 
it follows that for $p > 1/5$
\begin{align}\label{SqudW4}
\sum^5_{i=1}|\mathcal{C}_{A_iB_i}|=\frac{1+16\mathcal{N}(\rho_W(p))}{3}.
\end{align}
Thus the sum of PCCs is a linear function of Negativity and hence quantifies entanglement for the Werner state
(Eq. (\ref{wernerd})) for $d=4$.

\begin{widetext}
 \begin{center}
\begin{figure}[!t]
\includegraphics[scale=0.9]{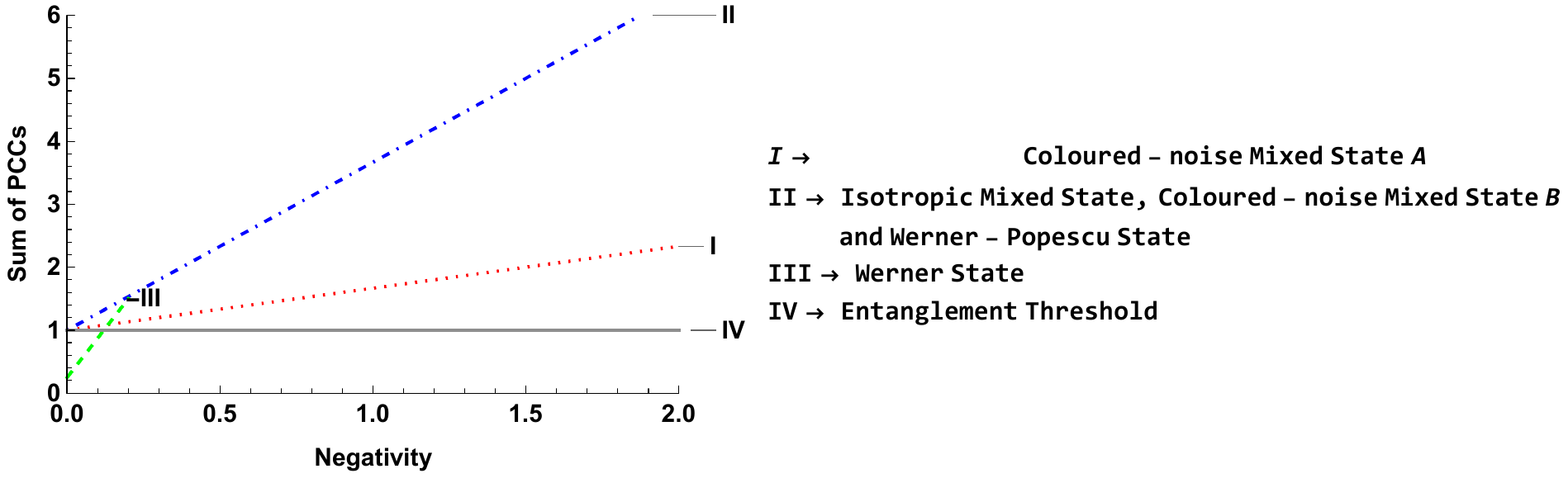}
\caption{\footnotesize  For $d=5$, the sum of PCCs is plotted as a function of negativity  for the  six  families of two-qudit states  indicated in the right hand side. 
The dotted line (I) corresponds to the sum of two PCCs versus negativity for the coloured-noise mixed state A given by Eq. (\ref{PCCccN5}) in the text.
The dot-dashed line (II) denotes the sum of six PCCs versus negativity for the isotropic mixed state, coloured-noise mixed state B and Werner-Popescu state given by Eqs. (\ref{Squd5I}) and (\ref{d5cnpccN}) and (\ref{d5WPNS}) respectively.
The dashed line (III) indicates the sum of six PCCs versus negativity for the Werner states given by Eq. (\ref{SqudW5}).
The horizontal line (IV) specifies entanglement threshold above which the states are entangled.}
\label{PlotD=5}
\end{figure}
\end{center}
\end{widetext}
\paragraph*{\bf{For d=5:}}
In order to certify entanglement of the Werner state given by Eq. (\ref{wernerd}) in dimension $d=5$, 
we use the sum of six PCCs $\sum^6_{i=1}|\mathcal{C}_{A_iB_i}|$, 
where $A_1=B_1=\sum_j a_j \ketbra{a_j}{a_j}$, $A_2=B_2=\sum_j b_j \ketbra{b_j}{b_j}$, 
$A_3=B_3=\sum_j e_j \ketbra{e_j}{e_j}$, $A_4=B_4=\sum_j g_j \ketbra{g_j}{g_j}$ and $A_5=B_5=\sum_j k_j \ketbra{k_j}{k_j}$.
For the six noncommuting  bases (which are not MUBs) given by Eq. (\ref{ncb5}) in Appendix \ref{NCB45} with the eigenvalues $a_0=b_0=e_0=g_0=k_0=l_0=+2$, $a_1=b_1=e_1=g_1=k_1=l_1=+1$, 
$a_2=b_2=e_2=g_2=k_2=l_2=0$, $a_3=b_3=e_3=g_3=k_3=l_3=-1$ and $a_4=b_4=e_4=g_4=k_4=l_4=-2$, 
the sum of six PCCs  is given as follows
 \begin{align}\label{sPw5}
 \sum^6_{i=1}|\mathcal{C}_{A_iB_i}|
 &=\frac{3}{2}|p|>1 \quad \text{iff} \quad  p>2/3.
\end{align}
Since  the Werner states given by Eq. (\ref{wernerd}) are entangled for $p>1/6$ in dimension $d=5$,
it follows that the  sum of six PCCs given by Eq. (\ref{sPw5}) being greater than $1$ provides sufficient criterion for the certification of entanglement
of the Werner states given by Eq. (\ref{wernerd}) in dimension $d=5$.  Interestingly, it is found that the expression for the sum of six PCCs obtained in Eq. (\ref{sPw5}) can also be obtained by the set of six MUBs given by Eq. (\ref{csmubd=5}) in Appendix \ref{MUBsd35}.
Next, we argue that the sum of PCCs given by Eq. (\ref{sPw5})  also provides quantification of certified entanglement of the Werner states.

For the two-qudit Werner state $\rho_W(p)$ given by Eq. (\ref{wernerd}) in dimension $d=5$, Negativity as defined in Ref. \cite{VW02}  computed from the partial transposed density matrix  is given by 
\begin{align}\label{nWqud5}
\mathcal{N}(\rho_W(p))=\max \Bigg\{ \dfrac{6p -1}{25} , 0 \Bigg \}	
\end{align}
which is nonzero if and only if $p>1/6$.  Now, note that using Eq. (\ref{nWqud5}), one can write for $p > 1/6$
\begin{align}\label{nWqud5'}
\mathcal{N}(\rho_W(p))= \dfrac{6p -1}{25}.
\end{align}
From the above Eq. (\ref{nWqud5'}), using Eq. (\ref{sPw5}) 
it follows that for $p > 1/6$
\begin{align}\label{SqudW5}
\sum^6_{i=1}|\mathcal{C}_{A_iB_i}|=\frac{1+25\mathcal{N}(\rho_W(p))}{4}.
\end{align}
Thus the sum of PCCs is a linear function of Negativity and hence quantifies entanglement for the Werner state (Eq. (\ref{wernerd})) for $d=5$.

\begin{table*}
\begin{tabular}{|c||c|c|c|}
\hline
&    Werner state in $d=3$     &      Werner state  in $d=4$ &      Werner state  in $d=5$    \\
\hline \hline
Range of Entanglement  &  $p > \frac{1}{4} $ & $p > \frac{1}{5} $ & $p > \frac{1}{6} $ \\
\hline
Entanglement certification  &&&
\\ by $d+1$ PCCs with noncommuting/MU bases  & $p  > \frac{1}{2} $  & $p > \frac{3}{5} $ & $p > \frac{2}{3} $ \\
\hline
Entanglement certification  based on &&& \\  
$d+1$ mutually unbiased measurements \cite{SLD15}  & $p  > \frac{1}{2} $  & $p > \frac{3}{5} $ & $p > \frac{2}{3} $ \\
\hline
\end{tabular}
\caption{The parameter ranges in which the Werner states for dimensions $d=3,4$ and $5$ are respectively entangled are given in the first row of the above Table.
The second and third rows show respectively the parameter ranges in which the 
entanglement of Werner states 
in  $d=3,4$ and $5$ are certified respectively using the  PCC based approach and by invoking mutually unbiased measurements \cite{SLD15}.}\label{}
\end{table*}

\subsection{Werner-Popescu states}
Here we address the entanglement characterization of the two-qudit Werner-Popescu states in the $d=4$ and $d=5$ cases
using PCCs, similar to the way discussed for the two-qutrit Werner-Popescu states.

\paragraph*{\bf{For d=4:}}

Now, to certify entanglement of the Werner-Popescu state given by Eq. (\ref{WPd}) in dimension $d=4$, 
we use the sum of five PCCs $\sum^5_{i=1}|\mathcal{C}_{A_iB_i}|$, 
where $A_1=B_1=\sum_j a_j \ketbra{a_j}{a_j}$, $A_2=B_2=\sum_j b_j \ketbra{b_j}{b_j}$, 
$A_3=B_3=\sum_j e_j \ketbra{e_j}{e_j}$, $A_4=B_4=\sum_j g_j \ketbra{g_j}{g_j}$ and $A_5=B_5=\sum_j k_j \ketbra{k_j}{k_j}$.
For the choice of five mutually unbiased bases given by Eq. (\ref{csmub4}) in Appendix \ref{NCB45}
with the eigenvalues $a_0=b_0=e_0=g_0=k_0=+2$, $a_1=b_1=e_1=g_1=k_1=+1$, $a_2=b_2=e_2=g_2=k_2=-1$ and $a_3=b_3=e_3=g_3=k_3=-2$, the sum of five PCCs is given by
 \begin{align}\label{d4WPNS}
 \sum^5_{i=1}|\mathcal{C}_{A_iB_i}|
 &=5p>1 \quad \text{iff} \quad  p>1/5. 
 \end{align}
Since the  Werner-Popescu state (given by Eq. (\ref{WPd}))  is entangled for $p>1/5$ for dimension $d=4$,
it follows that the  sum of five PCCs given by Eq. (\ref{d4WPNS}) being greater than $1$ provides necessary and sufficient criterion for the certification of entanglement
of the  Werner-Popescu state given by Eq. (\ref{WPd}) in dimension $d=4$.
Next, we argue that the sum of PCCs given by Eq. (\ref{d4WPNS})  also provides quantification of certified entanglement of the  Werner-Popescu state given by Eq. (\ref{WPd}) in dimension $d=4$.
 
For the  Werner-Popescu state  (given by Eq. (\ref{WPd})) in $d=4$, Negativity as defined in Ref. \cite{VW02} can be computed from the partial transposed density matrix
and is given by 
\begin{align}\label{nqud=4WP}
\mathcal{N}(\rho_{WP}(p))=\max \Bigg\{ \dfrac{3(5p -1)}{8} , 0 \Bigg \}		
\end{align}
which is nonzero if and only if $p>1/5$. 
Therefore, it follows that
Negativity of the Werner-Popescu state given by Eq. (\ref{WPd}) in $d=4$
provides the necessary and sufficient quantification of entanglement.
Note that using Eq. (\ref{nqud=4WP}), one can write Negativity of the entangled isotropic mixed state in $d=4$
as follows
\begin{align}\label{nqudWP4}
\mathcal{N}(\rho_{WP}(p))= \dfrac{3(5p -1)}{8}.
\end{align}
From the above Eq. (\ref{nqudWP4}), using Eq. (\ref{d4WPNS}) 
it follows that for $p > 1/5$
\begin{align}\label{Squd4WP}
\sum^5_{i=1}|\mathcal{C}_{A_iB_i}|=1+\frac{8}{3}\mathcal{N}(\rho_{WP}(p))
\end{align}
Hence the sum of PCCs is a linear function of Negativity and hence quantifies entanglement in this case.

\paragraph*{\bf{For d=5:}}
Similarly, 
now, to certify entanglement of the  Werner-Popescu state  given by Eq. (\ref{WPd}) in dimension $d=5$, 
we use the sum of six PCCs $\sum^6_{i=1}|\mathcal{C}_{A_iB_i}|$, 
where $A_1=B_1=\sum_j a_j \ketbra{a_j}{a_j}$, $A_2=B_2=\sum_j b_j \ketbra{b_j}{b_j}$, 
$A_3=B_3=\sum_j e_j \ketbra{e_j}{e_j}$, $A_4=B_4=\sum_j g_j \ketbra{g_j}{g_j}$, $A_5=B_5=\sum_j k_j \ketbra{k_j}{k_j}$
and $A_6=B_6=\sum_j l_j \ketbra{l_j}{l_j}$.
For the choice of six noncommuting bases which are not MUBs given by Eq. (\ref{ncb5}) in Appendix \ref{NCB45}
with the eigenvalues $a_0=b_0=e_0=g_0=k_0=l_0=+2$, $a_1=b_1=e_1=g_1=k_1=l_1=+1$, 
$a_2=b_2=e_2=g_2=k_2=l_2=0$, $a_3=b_3=e_3=g_3=k_3=l_3=-1$ and $a_4=b_4=e_4=g_4=k_4=l_4=-2$, the sum of six PCCs is given by
 \begin{align}\label{d5WPNS}
 \sum^6_{i=1}|\mathcal{C}_{A_iB_i}|
 &=6p>1 \quad \text{iff} \quad  p>1/6 
 \end{align}
Since the generalized Werner-Popescu state given by given by Eq. (\ref{WPd})  is entangled for $p>1/6$ for dimension $d=5$,
it follows that the  sum of six PCCs given by Eq. (\ref{d5WPNS}) being greater than $1$ provides necessary and sufficient criterion for the certification of entanglement
of the isotropic mixed state (\ref{WPd}) in dimension $d=5$.
Similar to the case $d=4$, we now argue that the sum of PCCs given by Eq. (\ref{d5WPNS})  also provides quantification of certified entanglement of the generalized Werner-Popescu state  (\ref{WPd}) in dimension $d=5$
in the following sense.

For the Werner-Popescu state (given by Eq. (\ref{WPd})) in $d=5$, Negativity as defined in Ref. \cite{VW02} 
is given by 
\begin{align}\label{nqud=5WP}
\mathcal{N}(\rho_{WP}(p))=\max \Bigg\{ \dfrac{2(6p -1)}{5} , 0 \Bigg \}		
\end{align}
which is nonzero if and only if $p>1/6$. 
Therefore, Negativity of the Werner-Popescu state in $d=5$
provides the necessary and sufficient quantification of entanglement.
Now, using Eq. (\ref{nqud=5WP}), Negativity of the entangled Werner-Popescu  state in $d=5$
is given by
\begin{align}\label{nqudWP5}
\mathcal{N}(\rho_{WP}(p))= \dfrac{2(6p -1)}{5}.
\end{align}
Using Eq. (\ref{d5WPNS}) 
it then follows that for $p > 1/6$
\begin{align}\label{Squd5WP}
\sum^5_{i=1}|\mathcal{C}_{A_iB_i}|=1+\frac{5}{2}\mathcal{N}(\rho_{WP}(p))
\end{align}
Thus, the sum of PCCs is a linear function of Negativity and hence quantifies entanglement in this case, too.

Note that Negativity of the Werner-Popescu state does not have a closed form of expression for arbitrary dimension $d$ as in the case of isotropic state.
Nevertheless, it is interesting that the relationship between the sum of $d+1$ PCCs and Negativity for the two-qudit Werner-Popescu state in the cases of $d=3,4$ and $5$ given by Eqs. (\ref{SqutWP}), (\ref{Squd4WP}) and (\ref{Squd5WP}) respectively has the same form as that for the two-qudit isotropic state in these cases given by Eqs. (\ref{SqutI}), (\ref{Squd4I}) and (\ref{Squd5I}) respectively.
\section{Concluding remarks}
In a nutshell, the work reported here demonstrates for dimensions $d=3,4$ and $5$ that the scheme formulated here  relating the experimentally measurable Pearson correlation coefficients (PCCs) with  Negativity as an entanglement measure is able to provide necessary and sufficient certification as well as quantification of entanglement  for a range of physically relevant mixed states such as isotropic  states, coloured-noise mixed states-A and Werner-Popescu states (see Figs. [\ref{PlotD=3},\ref{PlotD=4},\ref{PlotD=5}] illustrating the results).  Even for the Werner states in higher dimensions whose entanglement characterization has remained less explored by other approaches, the scheme discussed here in terms of PCCs is shown to furnish sufficient certification along with quantification of entanglement for dimensions $d=3,4$ and $5$ (also shown in Figs. [\ref{PlotD=3},\ref{PlotD=4},\ref{PlotD=5}]).  
Comparing the sufficient certification of entanglement for the Werner states using the PCC based approach with that provided by the entanglement certification procedure \cite{SLD15} based on d+1 mutually unbiased measurements, an interesting feature is noted  that the range of values of the mixedness parameter for which the Werner states for $d=3,4$ and $5$ are respectively certified to be entangled by both the approaches turn out to be the same (see Table $1$). However, the quantification of entanglement in these cases has remained unanalysed in terms  of the other approach \cite{SLD15}, while in our paper the PCC based approach is shown to be able to quantify entanglement of the Werner states for $d=3,4$ and $5$. Further, for the coloured-noise mixed states-B, we show that PCCs can be used for quantification of 
certified entanglement when Negativity is nonvanishing. 
Thus, the range of results obtained in this paper serve to reveal the strength of the PCC based approach and  provides  impetus for investigating its extension for  entanglement characterization in even higher dimensions than what has been considered in this work.

A key revelation of our treatment is that,  among different measures of entanglement in high dimensions, it is Negativity as the measure of entanglement which is found to be analytically and monotonically related to the quantitative measure of  correlations using combinations of PCCs in  noncommuting bases (which may or may not be mutually unbiased). On the other hand, for pure states in any dimension, it has been argued that it is the 
correlation in mutually unbiased bases as quantified by a suitable information-theoretic measure which is directly related to the 
entanglement of formation \cite{WMC+14,GW14}. The physical meaning of the latter as entanglement measure for the higher dimensional systems, interestingly, contrasts with that of Negativity. While entanglement of formation signifies the minimum
number of `ebits' required to prepare a given state using local operations and classical communication \cite{BDS+96, HHH+09},  Negativity, as mentioned earlier \cite{ES13}, can be regarded as an estimator of how many degrees of freedom of the subsystems are entangled, or, as determining the minimum number of
dimensions involved in the quantum correlation. These notions, thus, require a deeper holistic probing by taking into account the various theoretical studies on different entanglement measures \cite{EP99, Zyc99, MG04, MG'04, VAD01, CS08, Ero15} and the comparison between Negativity and entanglement of formation experimentally studied for the first time for higher dimensional system in the accompanying paper \cite{Sinha2019}.

\section*{Acknowledgement}
CJ acknowledges S. N. Bose Centre, Kolkata for the
postdoctoral fellowship and support from Ministry of Science and Technology of Taiwan(108-2811-M-006-501). The research of DH is supported by NASI Senior Scientist Fellowship. Thanks are due to Surya Narayan Banerjee (IISER Pune) for help in numerical computations. DH thanks Som Kanjilal for useful discussions.

\bibliography{JBS}

\begin{thebibliography}{80}
\expandafter\ifx\csname natexlab\endcsname\relax\def\natexlab#1{#1}\fi
\expandafter\ifx\csname bibnamefont\endcsname\relax
  \def\bibnamefont#1{#1}\fi
\expandafter\ifx\csname bibfnamefont\endcsname\relax
  \def\bibfnamefont#1{#1}\fi
\expandafter\ifx\csname citenamefont\endcsname\relax
  \def\citenamefont#1{#1}\fi
\expandafter\ifx\csname url\endcsname\relax
  \def\url#1{\texttt{#1}}\fi
\expandafter\ifx\csname urlprefix\endcsname\relax\def\urlprefix{URL }\fi
\providecommand{\bibinfo}[2]{#2}
\providecommand{\eprint}[2][]{\url{#2}}

\bibitem[{\citenamefont{Ekert}(1991)}]{ek}
\bibinfo{author}{\bibfnamefont{A.~K.} \bibnamefont{Ekert}},
  \bibinfo{journal}{Phys. Rev. Lett.} \textbf{\bibinfo{volume}{67}},
  \bibinfo{pages}{661} (\bibinfo{year}{1991}),
  \urlprefix\url{https://link.aps.org/doi/10.1103/PhysRevLett.67.661}.

\bibitem[{\citenamefont{Bennett and Wiesner}(1992)}]{ben}
\bibinfo{author}{\bibfnamefont{C.~H.} \bibnamefont{Bennett}} \bibnamefont{and}
  \bibinfo{author}{\bibfnamefont{S.~J.} \bibnamefont{Wiesner}},
  \bibinfo{journal}{Phys. Rev. Lett.} \textbf{\bibinfo{volume}{69}},
  \bibinfo{pages}{2881} (\bibinfo{year}{1992}),
  \urlprefix\url{https://link.aps.org/doi/10.1103/PhysRevLett.69.2881}.

\bibitem[{\citenamefont{Bennett et~al.}(1993)\citenamefont{Bennett, Brassard,
  Cr\'epeau, Jozsa, Peres, and Wootters}}]{benn}
\bibinfo{author}{\bibfnamefont{C.~H.} \bibnamefont{Bennett}},
  \bibinfo{author}{\bibfnamefont{G.}~\bibnamefont{Brassard}},
  \bibinfo{author}{\bibfnamefont{C.}~\bibnamefont{Cr\'epeau}},
  \bibinfo{author}{\bibfnamefont{R.}~\bibnamefont{Jozsa}},
  \bibinfo{author}{\bibfnamefont{A.}~\bibnamefont{Peres}}, \bibnamefont{and}
  \bibinfo{author}{\bibfnamefont{W.~K.} \bibnamefont{Wootters}},
  \bibinfo{journal}{Phys. Rev. Lett.} \textbf{\bibinfo{volume}{70}},
  \bibinfo{pages}{1895} (\bibinfo{year}{1993}),
  \urlprefix\url{https://link.aps.org/doi/10.1103/PhysRevLett.70.1895}.

\bibitem[{\citenamefont{Ac\'{\i}n et~al.}(2007)\citenamefont{Ac\'{\i}n,
  Brunner, Gisin, Massar, Pironio, and Scarani}}]{ABG+07}
\bibinfo{author}{\bibfnamefont{A.}~\bibnamefont{Ac\'{\i}n}},
  \bibinfo{author}{\bibfnamefont{N.}~\bibnamefont{Brunner}},
  \bibinfo{author}{\bibfnamefont{N.}~\bibnamefont{Gisin}},
  \bibinfo{author}{\bibfnamefont{S.}~\bibnamefont{Massar}},
  \bibinfo{author}{\bibfnamefont{S.}~\bibnamefont{Pironio}}, \bibnamefont{and}
  \bibinfo{author}{\bibfnamefont{V.}~\bibnamefont{Scarani}},
  \bibinfo{journal}{Phys. Rev. Lett.} \textbf{\bibinfo{volume}{98}},
  \bibinfo{pages}{230501} (\bibinfo{year}{2007}),
  \urlprefix\url{https://link.aps.org/doi/10.1103/PhysRevLett.98.230501}.

\bibitem[{\citenamefont{Jozsa and Linden}(2003)}]{JL03}
\bibinfo{author}{\bibfnamefont{R.}~\bibnamefont{Jozsa}} \bibnamefont{and}
  \bibinfo{author}{\bibfnamefont{N.}~\bibnamefont{Linden}},
  \bibinfo{journal}{Proc. Roy. Soc. A} \textbf{\bibinfo{volume}{459}},
  \bibinfo{pages}{2011} (\bibinfo{year}{2003}), ISSN \bibinfo{issn}{1364-5021},
  \urlprefix\url{http://rspa.royalsocietypublishing.org/content/459/2036/2011}.

\bibitem[{\citenamefont{Brukner et~al.}(2004)\citenamefont{Brukner,
  \.{Z}ukowski, Pan, and Zeilinger}}]{bruk}
\bibinfo{author}{\bibfnamefont{C.}~\bibnamefont{Brukner}},
  \bibinfo{author}{\bibfnamefont{M.}~\bibnamefont{\.{Z}ukowski}},
  \bibinfo{author}{\bibfnamefont{J.-W.} \bibnamefont{Pan}}, \bibnamefont{and}
  \bibinfo{author}{\bibfnamefont{A.}~\bibnamefont{Zeilinger}},
  \bibinfo{journal}{Phys. Rev. Lett.} \textbf{\bibinfo{volume}{92}},
  \bibinfo{pages}{127901} (\bibinfo{year}{2004}),
  \urlprefix\url{https://link.aps.org/doi/10.1103/PhysRevLett.92.127901}.

\bibitem[{\citenamefont{Buhrman et~al.}(2010)\citenamefont{Buhrman, Cleve,
  Massar, and de~Wolf}}]{BCM+10}
\bibinfo{author}{\bibfnamefont{H.}~\bibnamefont{Buhrman}},
  \bibinfo{author}{\bibfnamefont{R.}~\bibnamefont{Cleve}},
  \bibinfo{author}{\bibfnamefont{S.}~\bibnamefont{Massar}}, \bibnamefont{and}
  \bibinfo{author}{\bibfnamefont{R.}~\bibnamefont{de~Wolf}},
  \bibinfo{journal}{Rev. Mod. Phys.} \textbf{\bibinfo{volume}{82}},
  \bibinfo{pages}{665} (\bibinfo{year}{2010}),
  \urlprefix\url{https://link.aps.org/doi/10.1103/RevModPhys.82.665}.

\bibitem[{\citenamefont{Pironio et~al.}(2010)\citenamefont{Pironio, Acin,
  Massar, de~la Giroday, Matsukevich, Maunz, Olmschenk, Hayes, Luo, Manning
  et~al.}}]{PAM+10}
\bibinfo{author}{\bibfnamefont{S.}~\bibnamefont{Pironio}},
  \bibinfo{author}{\bibfnamefont{A.}~\bibnamefont{Acin}},
  \bibinfo{author}{\bibfnamefont{S.}~\bibnamefont{Massar}},
  \bibinfo{author}{\bibfnamefont{A.~B.} \bibnamefont{de~la Giroday}},
  \bibinfo{author}{\bibfnamefont{D.~N.} \bibnamefont{Matsukevich}},
  \bibinfo{author}{\bibfnamefont{P.}~\bibnamefont{Maunz}},
  \bibinfo{author}{\bibfnamefont{S.}~\bibnamefont{Olmschenk}},
  \bibinfo{author}{\bibfnamefont{D.}~\bibnamefont{Hayes}},
  \bibinfo{author}{\bibfnamefont{L.}~\bibnamefont{Luo}},
  \bibinfo{author}{\bibfnamefont{T.~A.} \bibnamefont{Manning}},
  \bibnamefont{et~al.}, \bibinfo{journal}{Nat Phys}
  \textbf{\bibinfo{volume}{464}}, \bibinfo{pages}{1021} (\bibinfo{year}{2010}).

\bibitem[{\citenamefont{Nieto-Silleras
  et~al.}(2014)\citenamefont{Nieto-Silleras, Pironio, and Silman}}]{NPS14}
\bibinfo{author}{\bibfnamefont{O.}~\bibnamefont{Nieto-Silleras}},
  \bibinfo{author}{\bibfnamefont{S.}~\bibnamefont{Pironio}}, \bibnamefont{and}
  \bibinfo{author}{\bibfnamefont{J.}~\bibnamefont{Silman}},
  \bibinfo{journal}{New Journal of Physics} \textbf{\bibinfo{volume}{16}},
  \bibinfo{pages}{013035} (\bibinfo{year}{2014}),
  \urlprefix\url{http://stacks.iop.org/1367-2630/16/i=1/a=013035}.

\bibitem[{\citenamefont{Bechmann-Pasquinucci and Tittel}(2000)}]{bech}
\bibinfo{author}{\bibfnamefont{H.}~\bibnamefont{Bechmann-Pasquinucci}}
  \bibnamefont{and} \bibinfo{author}{\bibfnamefont{W.}~\bibnamefont{Tittel}},
  \bibinfo{journal}{Phys. Rev. A} \textbf{\bibinfo{volume}{61}},
  \bibinfo{pages}{062308} (\bibinfo{year}{2000}),
  \urlprefix\url{https://link.aps.org/doi/10.1103/PhysRevA.61.062308}.

\bibitem[{\citenamefont{Cerf et~al.}(2002)\citenamefont{Cerf, Bourennane,
  Karlsson, and Gisin}}]{CBK02}
\bibinfo{author}{\bibfnamefont{N.~J.} \bibnamefont{Cerf}},
  \bibinfo{author}{\bibfnamefont{M.}~\bibnamefont{Bourennane}},
  \bibinfo{author}{\bibfnamefont{A.}~\bibnamefont{Karlsson}}, \bibnamefont{and}
  \bibinfo{author}{\bibfnamefont{N.}~\bibnamefont{Gisin}},
  \bibinfo{journal}{Phys. Rev. Lett.} \textbf{\bibinfo{volume}{88}},
  \bibinfo{pages}{127902} (\bibinfo{year}{2002}),
  \urlprefix\url{https://link.aps.org/doi/10.1103/PhysRevLett.88.127902}.

\bibitem[{\citenamefont{Bru\ss{} et~al.}(2003)\citenamefont{Bru\ss{},
  Christandl, Ekert, Englert, Kaszlikowski, and Macchiavello}}]{BCE+03}
\bibinfo{author}{\bibfnamefont{D.}~\bibnamefont{Bru\ss{}}},
  \bibinfo{author}{\bibfnamefont{M.}~\bibnamefont{Christandl}},
  \bibinfo{author}{\bibfnamefont{A.}~\bibnamefont{Ekert}},
  \bibinfo{author}{\bibfnamefont{B.-G.} \bibnamefont{Englert}},
  \bibinfo{author}{\bibfnamefont{D.}~\bibnamefont{Kaszlikowski}},
  \bibnamefont{and}
  \bibinfo{author}{\bibfnamefont{C.}~\bibnamefont{Macchiavello}},
  \bibinfo{journal}{Phys. Rev. Lett.} \textbf{\bibinfo{volume}{91}},
  \bibinfo{pages}{097901} (\bibinfo{year}{2003}),
  \urlprefix\url{https://link.aps.org/doi/10.1103/PhysRevLett.91.097901}.

\bibitem[{\citenamefont{Sheridan and Scarani}(2010)}]{SV10}
\bibinfo{author}{\bibfnamefont{L.}~\bibnamefont{Sheridan}} \bibnamefont{and}
  \bibinfo{author}{\bibfnamefont{V.}~\bibnamefont{Scarani}},
  \bibinfo{journal}{Phys. Rev. A} \textbf{\bibinfo{volume}{82}},
  \bibinfo{pages}{030301} (\bibinfo{year}{2010}),
  \urlprefix\url{https://link.aps.org/doi/10.1103/PhysRevA.82.030301}.

\bibitem[{\citenamefont{Huber and Pawlowski}(2013)}]{hub}
\bibinfo{author}{\bibfnamefont{M.}~\bibnamefont{Huber}} \bibnamefont{and}
  \bibinfo{author}{\bibfnamefont{M.}~\bibnamefont{Pawlowski}},
  \bibinfo{journal}{Phys. Rev. A} \textbf{\bibinfo{volume}{88}},
  \bibinfo{pages}{032309} (\bibinfo{year}{2013}),
  \urlprefix\url{https://link.aps.org/doi/10.1103/PhysRevA.88.032309}.

\bibitem[{\citenamefont{Bennett et~al.}(1999)\citenamefont{Bennett, Shor,
  Smolin, and Thapliyal}}]{bennett}
\bibinfo{author}{\bibfnamefont{C.~H.} \bibnamefont{Bennett}},
  \bibinfo{author}{\bibfnamefont{P.~W.} \bibnamefont{Shor}},
  \bibinfo{author}{\bibfnamefont{J.~A.} \bibnamefont{Smolin}},
  \bibnamefont{and} \bibinfo{author}{\bibfnamefont{A.~V.}
  \bibnamefont{Thapliyal}}, \bibinfo{journal}{Phys. Rev. Lett.}
  \textbf{\bibinfo{volume}{83}}, \bibinfo{pages}{3081} (\bibinfo{year}{1999}),
  \urlprefix\url{https://link.aps.org/doi/10.1103/PhysRevLett.83.3081}.

\bibitem[{\citenamefont{Wang et~al.}(2005)\citenamefont{Wang, Deng, Li, Liu,
  and Long}}]{WDF+05}
\bibinfo{author}{\bibfnamefont{C.}~\bibnamefont{Wang}},
  \bibinfo{author}{\bibfnamefont{F.-G.} \bibnamefont{Deng}},
  \bibinfo{author}{\bibfnamefont{Y.-S.} \bibnamefont{Li}},
  \bibinfo{author}{\bibfnamefont{X.-S.} \bibnamefont{Liu}}, \bibnamefont{and}
  \bibinfo{author}{\bibfnamefont{G.~L.} \bibnamefont{Long}},
  \bibinfo{journal}{Phys. Rev. A} \textbf{\bibinfo{volume}{71}},
  \bibinfo{pages}{044305} (\bibinfo{year}{2005}),
  \urlprefix\url{https://link.aps.org/doi/10.1103/PhysRevA.71.044305}.

\bibitem[{\citenamefont{Brunner and Roux}(2013)}]{brun}
\bibinfo{author}{\bibfnamefont{T.}~\bibnamefont{Brunner}} \bibnamefont{and}
  \bibinfo{author}{\bibfnamefont{F.~S.} \bibnamefont{Roux}},
  \bibinfo{journal}{New Journal of Physics} \textbf{\bibinfo{volume}{15}},
  \bibinfo{pages}{063005} (\bibinfo{year}{2013}),
  \urlprefix\url{http://stacks.iop.org/1367-2630/15/i=6/a=063005}.

\bibitem[{\citenamefont{V\'ertesi et~al.}(2010)\citenamefont{V\'ertesi,
  Pironio, and Brunner}}]{ver}
\bibinfo{author}{\bibfnamefont{T.}~\bibnamefont{V\'ertesi}},
  \bibinfo{author}{\bibfnamefont{S.}~\bibnamefont{Pironio}}, \bibnamefont{and}
  \bibinfo{author}{\bibfnamefont{N.}~\bibnamefont{Brunner}},
  \bibinfo{journal}{Phys. Rev. Lett.} \textbf{\bibinfo{volume}{104}},
  \bibinfo{pages}{060401} (\bibinfo{year}{2010}),
  \urlprefix\url{https://link.aps.org/doi/10.1103/PhysRevLett.104.060401}.

\bibitem[{\citenamefont{Ioannou}(2007)}]{IOA17}
\bibinfo{author}{\bibfnamefont{L.~M.} \bibnamefont{Ioannou}},
  \bibinfo{journal}{Quantum Information and Computation}
  \textbf{\bibinfo{volume}{7}}, \bibinfo{pages}{335} (\bibinfo{year}{2007}).

\bibitem[{\citenamefont{Shahandeh et~al.}(2013)\citenamefont{Shahandeh,
  Sperling, and Vogel}}]{SSV13}
\bibinfo{author}{\bibfnamefont{F.}~\bibnamefont{Shahandeh}},
  \bibinfo{author}{\bibfnamefont{J.}~\bibnamefont{Sperling}}, \bibnamefont{and}
  \bibinfo{author}{\bibfnamefont{W.}~\bibnamefont{Vogel}},
  \bibinfo{journal}{Phys. Rev. A} \textbf{\bibinfo{volume}{88}},
  \bibinfo{pages}{062323} (\bibinfo{year}{2013}),
  \urlprefix\url{https://link.aps.org/doi/10.1103/PhysRevA.88.062323}.

\bibitem[{\citenamefont{Shahandeh et~al.}(2014)\citenamefont{Shahandeh,
  Sperling, and Vogel}}]{SSV14}
\bibinfo{author}{\bibfnamefont{F.}~\bibnamefont{Shahandeh}},
  \bibinfo{author}{\bibfnamefont{J.}~\bibnamefont{Sperling}}, \bibnamefont{and}
  \bibinfo{author}{\bibfnamefont{W.}~\bibnamefont{Vogel}},
  \bibinfo{journal}{Phys. Rev. Lett.} \textbf{\bibinfo{volume}{113}},
  \bibinfo{pages}{260502} (\bibinfo{year}{2014}),
  \urlprefix\url{https://link.aps.org/doi/10.1103/PhysRevLett.113.260502}.

\bibitem[{\citenamefont{Shahandeh et~al.}(2017)\citenamefont{Shahandeh, Hall,
  and Ralph}}]{SHR17}
\bibinfo{author}{\bibfnamefont{F.}~\bibnamefont{Shahandeh}},
  \bibinfo{author}{\bibfnamefont{M.~J.~W.} \bibnamefont{Hall}},
  \bibnamefont{and} \bibinfo{author}{\bibfnamefont{T.~C.} \bibnamefont{Ralph}},
  \bibinfo{journal}{Phys. Rev. Lett.} \textbf{\bibinfo{volume}{118}},
  \bibinfo{pages}{150505} (\bibinfo{year}{2017}),
  \urlprefix\url{https://link.aps.org/doi/10.1103/PhysRevLett.118.150505}.

\bibitem[{\citenamefont{\ifmmode \check{S}\else
  \v{S}\fi{}upi\ifmmode~\acute{c}\else \'{c}\fi{}
  et~al.}(2017)\citenamefont{\ifmmode \check{S}\else
  \v{S}\fi{}upi\ifmmode~\acute{c}\else \'{c}\fi{}, Skrzypczyk, and
  Cavalcanti}}]{SSC17}
\bibinfo{author}{\bibfnamefont{I.}~\bibnamefont{\ifmmode \check{S}\else
  \v{S}\fi{}upi\ifmmode~\acute{c}\else \'{c}\fi{}}},
  \bibinfo{author}{\bibfnamefont{P.}~\bibnamefont{Skrzypczyk}},
  \bibnamefont{and}
  \bibinfo{author}{\bibfnamefont{D.}~\bibnamefont{Cavalcanti}},
  \bibinfo{journal}{Phys. Rev. A} \textbf{\bibinfo{volume}{95}},
  \bibinfo{pages}{042340} (\bibinfo{year}{2017}),
  \urlprefix\url{https://link.aps.org/doi/10.1103/PhysRevA.95.042340}.

\bibitem[{\citenamefont{Tiranov et~al.}(2017)\citenamefont{Tiranov, Designolle,
  Cruzeiro, Lavoie, Brunner, Afzelius, Huber, and Gisin}}]{TDC+17}
\bibinfo{author}{\bibfnamefont{A.}~\bibnamefont{Tiranov}},
  \bibinfo{author}{\bibfnamefont{S.}~\bibnamefont{Designolle}},
  \bibinfo{author}{\bibfnamefont{E.~Z.} \bibnamefont{Cruzeiro}},
  \bibinfo{author}{\bibfnamefont{J.}~\bibnamefont{Lavoie}},
  \bibinfo{author}{\bibfnamefont{N.}~\bibnamefont{Brunner}},
  \bibinfo{author}{\bibfnamefont{M.}~\bibnamefont{Afzelius}},
  \bibinfo{author}{\bibfnamefont{M.}~\bibnamefont{Huber}}, \bibnamefont{and}
  \bibinfo{author}{\bibfnamefont{N.}~\bibnamefont{Gisin}},
  \bibinfo{journal}{Phys. Rev. A} \textbf{\bibinfo{volume}{96}},
  \bibinfo{pages}{040303} (\bibinfo{year}{2017}),
  \urlprefix\url{https://link.aps.org/doi/10.1103/PhysRevA.96.040303}.

\bibitem[{\citenamefont{Martin et~al.}(2017)\citenamefont{Martin, Guerreiro,
  Tiranov, Designolle, Fr\"owis, Brunner, Huber, and Gisin}}]{MGT+17}
\bibinfo{author}{\bibfnamefont{A.}~\bibnamefont{Martin}},
  \bibinfo{author}{\bibfnamefont{T.}~\bibnamefont{Guerreiro}},
  \bibinfo{author}{\bibfnamefont{A.}~\bibnamefont{Tiranov}},
  \bibinfo{author}{\bibfnamefont{S.}~\bibnamefont{Designolle}},
  \bibinfo{author}{\bibfnamefont{F.}~\bibnamefont{Fr\"owis}},
  \bibinfo{author}{\bibfnamefont{N.}~\bibnamefont{Brunner}},
  \bibinfo{author}{\bibfnamefont{M.}~\bibnamefont{Huber}}, \bibnamefont{and}
  \bibinfo{author}{\bibfnamefont{N.}~\bibnamefont{Gisin}},
  \bibinfo{journal}{Phys. Rev. Lett.} \textbf{\bibinfo{volume}{118}},
  \bibinfo{pages}{110501} (\bibinfo{year}{2017}),
  \urlprefix\url{https://link.aps.org/doi/10.1103/PhysRevLett.118.110501}.

\bibitem[{\citenamefont{Bavaresco et~al.}(2018)\citenamefont{Bavaresco,
  Natalia, Klöckl, Pivoluska, Erker, Friis, Malik, and Huber}}]{BVK+17}
\bibinfo{author}{\bibfnamefont{J.}~\bibnamefont{Bavaresco}},
  \bibinfo{author}{\bibfnamefont{H.~V.} \bibnamefont{Natalia}},
  \bibinfo{author}{\bibfnamefont{C.}~\bibnamefont{Klöckl}},
  \bibinfo{author}{\bibfnamefont{M.}~\bibnamefont{Pivoluska}},
  \bibinfo{author}{\bibfnamefont{P.}~\bibnamefont{Erker}},
  \bibinfo{author}{\bibfnamefont{N.}~\bibnamefont{Friis}},
  \bibinfo{author}{\bibfnamefont{M.}~\bibnamefont{Malik}}, \bibnamefont{and}
  \bibinfo{author}{\bibfnamefont{M.}~\bibnamefont{Huber}},
  \bibinfo{journal}{Nature Physics} \textbf{\bibinfo{volume}{14}},
  \bibinfo{pages}{1032} (\bibinfo{year}{2018}).

\bibitem[{\citenamefont{Schneeloch and Howland}(2018)}]{SH18}
\bibinfo{author}{\bibfnamefont{J.}~\bibnamefont{Schneeloch}} \bibnamefont{and}
  \bibinfo{author}{\bibfnamefont{G.~A.} \bibnamefont{Howland}},
  \bibinfo{journal}{Phys. Rev. A} \textbf{\bibinfo{volume}{97}},
  \bibinfo{pages}{042338} (\bibinfo{year}{2018}),
  \urlprefix\url{https://link.aps.org/doi/10.1103/PhysRevA.97.042338}.

\bibitem[{\citenamefont{Roy}(2005)}]{Roy05}
\bibinfo{author}{\bibfnamefont{S.~M.} \bibnamefont{Roy}},
  \bibinfo{journal}{Phys. Rev. Lett.} \textbf{\bibinfo{volume}{94}},
  \bibinfo{pages}{010402} (\bibinfo{year}{2005}),
  \urlprefix\url{https://link.aps.org/doi/10.1103/PhysRevLett.94.010402}.

\bibitem[{\citenamefont{Datta et~al.}(2017)\citenamefont{Datta, Agrawal, and
  Choudhary}}]{DAC17}
\bibinfo{author}{\bibfnamefont{C.}~\bibnamefont{Datta}},
  \bibinfo{author}{\bibfnamefont{P.}~\bibnamefont{Agrawal}}, \bibnamefont{and}
  \bibinfo{author}{\bibfnamefont{S.~K.} \bibnamefont{Choudhary}},
  \bibinfo{journal}{Phys. Rev. A} \textbf{\bibinfo{volume}{95}},
  \bibinfo{pages}{042323} (\bibinfo{year}{2017}),
  \urlprefix\url{https://link.aps.org/doi/10.1103/PhysRevA.95.042323}.

\bibitem[{\citenamefont{G\"uhne and L\"utkenhaus}(2006)}]{GL06}
\bibinfo{author}{\bibfnamefont{O.}~\bibnamefont{G\"uhne}} \bibnamefont{and}
  \bibinfo{author}{\bibfnamefont{N.}~\bibnamefont{L\"utkenhaus}},
  \bibinfo{journal}{Phys. Rev. Lett.} \textbf{\bibinfo{volume}{96}},
  \bibinfo{pages}{170502} (\bibinfo{year}{2006}),
  \urlprefix\url{https://link.aps.org/doi/10.1103/PhysRevLett.96.170502}.

\bibitem[{\citenamefont{Arrazola et~al.}(2012)\citenamefont{Arrazola,
  Gittsovich, and L\"utkenhaus}}]{AGL12}
\bibinfo{author}{\bibfnamefont{J.~M.} \bibnamefont{Arrazola}},
  \bibinfo{author}{\bibfnamefont{O.}~\bibnamefont{Gittsovich}},
  \bibnamefont{and}
  \bibinfo{author}{\bibfnamefont{N.}~\bibnamefont{L\"utkenhaus}},
  \bibinfo{journal}{Phys. Rev. A} \textbf{\bibinfo{volume}{85}},
  \bibinfo{pages}{062327} (\bibinfo{year}{2012}),
  \urlprefix\url{https://link.aps.org/doi/10.1103/PhysRevA.85.062327}.

\bibitem[{\citenamefont{Huang et~al.}(2016)\citenamefont{Huang, Maccone, Karim,
  Macchiavello, Chapman, and Peruzzo}}]{HMK+16}
\bibinfo{author}{\bibfnamefont{Z.}~\bibnamefont{Huang}},
  \bibinfo{author}{\bibfnamefont{L.}~\bibnamefont{Maccone}},
  \bibinfo{author}{\bibfnamefont{A.}~\bibnamefont{Karim}},
  \bibinfo{author}{\bibfnamefont{C.}~\bibnamefont{Macchiavello}},
  \bibinfo{author}{\bibfnamefont{R.~J.} \bibnamefont{Chapman}},
  \bibnamefont{and} \bibinfo{author}{\bibfnamefont{A.}~\bibnamefont{Peruzzo}},
  \bibinfo{journal}{Sci. Rep} \textbf{\bibinfo{volume}{6}},
  \bibinfo{pages}{27637} (\bibinfo{year}{2016}).

\bibitem[{\citenamefont{Maccone et~al.}(2015)\citenamefont{Maccone, Bru\ss{},
  and Macchiavello}}]{MBM15}
\bibinfo{author}{\bibfnamefont{L.}~\bibnamefont{Maccone}},
  \bibinfo{author}{\bibfnamefont{D.}~\bibnamefont{Bru\ss{}}}, \bibnamefont{and}
  \bibinfo{author}{\bibfnamefont{C.}~\bibnamefont{Macchiavello}},
  \bibinfo{journal}{Phys. Rev. Lett.} \textbf{\bibinfo{volume}{114}},
  \bibinfo{pages}{130401} (\bibinfo{year}{2015}),
  \urlprefix\url{https://link.aps.org/doi/10.1103/PhysRevLett.114.130401}.

\bibitem[{\citenamefont{Spengler et~al.}(2012)\citenamefont{Spengler, Huber,
  Brierley, Adaktylos, and Hiesmayr}}]{SHB+12}
\bibinfo{author}{\bibfnamefont{C.}~\bibnamefont{Spengler}},
  \bibinfo{author}{\bibfnamefont{M.}~\bibnamefont{Huber}},
  \bibinfo{author}{\bibfnamefont{S.}~\bibnamefont{Brierley}},
  \bibinfo{author}{\bibfnamefont{T.}~\bibnamefont{Adaktylos}},
  \bibnamefont{and} \bibinfo{author}{\bibfnamefont{B.~C.}
  \bibnamefont{Hiesmayr}}, \bibinfo{journal}{Phys. Rev. A}
  \textbf{\bibinfo{volume}{86}}, \bibinfo{pages}{022311}
  (\bibinfo{year}{2012}),
  \urlprefix\url{https://link.aps.org/doi/10.1103/PhysRevA.86.022311}.

\bibitem[{\citenamefont{Paul~Erker}(2017)}]{EKH17}
\bibinfo{author}{\bibfnamefont{M.~H.} \bibnamefont{Paul~Erker},
  \bibfnamefont{Mario~Krenn}}, \bibinfo{journal}{Quantum}
  \textbf{\bibinfo{volume}{1}}, \bibinfo{pages}{22} (\bibinfo{year}{2017}),
  \urlprefix\url{https://doi.org/10.22331/q-2017-07-28-22}.

\bibitem[{\citenamefont{Pearson}(1895)}]{pea}
\bibinfo{author}{\bibfnamefont{K.}~\bibnamefont{Pearson}},
  \bibinfo{journal}{Proc. Roy. Soc.(London)} \textbf{\bibinfo{volume}{58}},
  \bibinfo{pages}{240} (\bibinfo{year}{1895}).

\bibitem[{\citenamefont{Boccaletti et~al.}(2002)\citenamefont{Boccaletti,
  Kurths, Osipov, Valladares, and Zhou}}]{BKO02}
\bibinfo{author}{\bibfnamefont{S.}~\bibnamefont{Boccaletti}},
  \bibinfo{author}{\bibfnamefont{J.}~\bibnamefont{Kurths}},
  \bibinfo{author}{\bibfnamefont{G.}~\bibnamefont{Osipov}},
  \bibinfo{author}{\bibfnamefont{D.}~\bibnamefont{Valladares}},
  \bibnamefont{and} \bibinfo{author}{\bibfnamefont{C.}~\bibnamefont{Zhou}},
  \bibinfo{journal}{Physics Reports} \textbf{\bibinfo{volume}{366}},
  \bibinfo{pages}{1 } (\bibinfo{year}{2002}), ISSN \bibinfo{issn}{0370-1573}.

\bibitem[{\citenamefont{Bellomo et~al.}(2017)\citenamefont{Bellomo, Giorgi,
  Palma, and Zambrini}}]{BGP+17}
\bibinfo{author}{\bibfnamefont{B.}~\bibnamefont{Bellomo}},
  \bibinfo{author}{\bibfnamefont{G.~L.} \bibnamefont{Giorgi}},
  \bibinfo{author}{\bibfnamefont{G.~M.} \bibnamefont{Palma}}, \bibnamefont{and}
  \bibinfo{author}{\bibfnamefont{R.}~\bibnamefont{Zambrini}},
  \bibinfo{journal}{Phys. Rev. A} \textbf{\bibinfo{volume}{95}},
  \bibinfo{pages}{043807} (\bibinfo{year}{2017}),
  \urlprefix\url{https://link.aps.org/doi/10.1103/PhysRevA.95.043807}.

\bibitem[{\citenamefont{Pozsgay et~al.}(2017)\citenamefont{Pozsgay, Hirsch,
  Branciard, and Brunner}}]{PHB+17}
\bibinfo{author}{\bibfnamefont{V.}~\bibnamefont{Pozsgay}},
  \bibinfo{author}{\bibfnamefont{F.}~\bibnamefont{Hirsch}},
  \bibinfo{author}{\bibfnamefont{C.}~\bibnamefont{Branciard}},
  \bibnamefont{and} \bibinfo{author}{\bibfnamefont{N.}~\bibnamefont{Brunner}},
  \bibinfo{journal}{Phys. Rev. A} \textbf{\bibinfo{volume}{96}},
  \bibinfo{pages}{062128} (\bibinfo{year}{2017}),
  \urlprefix\url{https://link.aps.org/doi/10.1103/PhysRevA.96.062128}.

\bibitem[{\citenamefont{Ghosh et~al.}(2018)\citenamefont{Ghosh, Jennewein,
  Kolenderski, and Sinha}}]{GJK+17}
\bibinfo{author}{\bibfnamefont{D.}~\bibnamefont{Ghosh}},
  \bibinfo{author}{\bibfnamefont{T.}~\bibnamefont{Jennewein}},
  \bibinfo{author}{\bibfnamefont{P.}~\bibnamefont{Kolenderski}},
  \bibnamefont{and} \bibinfo{author}{\bibfnamefont{U.}~\bibnamefont{Sinha}},
  \bibinfo{journal}{OSA Continuum} \textbf{\bibinfo{volume}{1}},
  \bibinfo{pages}{996} (\bibinfo{year}{2018}),
  \urlprefix\url{http://www.osapublishing.org/osac/abstract.cfm?URI=osac-1-3-996}.

\bibitem[{\citenamefont{Ghosh et~al.}(2019)\citenamefont{Ghosh, Jennewein, and
  Sinha}}]{Sinha2019}
\bibinfo{author}{\bibfnamefont{D.}~\bibnamefont{Ghosh}},
  \bibinfo{author}{\bibfnamefont{T.}~\bibnamefont{Jennewein}},
  \bibnamefont{and} \bibinfo{author}{\bibfnamefont{U.}~\bibnamefont{Sinha}},
  \bibinfo{journal}{Accompanying manuscript}  (\bibinfo{year}{2019}).

\bibitem[{\citenamefont{Durt et~al.}(2003)\citenamefont{Durt, Cerf, Gisin, and
  \ifmmode~\dot{Z}\else \.{Z}\fi{}ukowski}}]{DCG+03}
\bibinfo{author}{\bibfnamefont{T.}~\bibnamefont{Durt}},
  \bibinfo{author}{\bibfnamefont{N.~J.} \bibnamefont{Cerf}},
  \bibinfo{author}{\bibfnamefont{N.}~\bibnamefont{Gisin}}, \bibnamefont{and}
  \bibinfo{author}{\bibfnamefont{M.}~\bibnamefont{\ifmmode~\dot{Z}\else
  \.{Z}\fi{}ukowski}}, \bibinfo{journal}{Phys. Rev. A}
  \textbf{\bibinfo{volume}{67}}, \bibinfo{pages}{012311}
  (\bibinfo{year}{2003}),
  \urlprefix\url{https://link.aps.org/doi/10.1103/PhysRevA.67.012311}.

\bibitem[{\citenamefont{Greentree et~al.}(2004)\citenamefont{Greentree,
  Schirmer, Green, Hollenberg, Hamilton, and Clark}}]{green}
\bibinfo{author}{\bibfnamefont{A.~D.} \bibnamefont{Greentree}},
  \bibinfo{author}{\bibfnamefont{S.~G.} \bibnamefont{Schirmer}},
  \bibinfo{author}{\bibfnamefont{F.}~\bibnamefont{Green}},
  \bibinfo{author}{\bibfnamefont{L.~C.~L.} \bibnamefont{Hollenberg}},
  \bibinfo{author}{\bibfnamefont{A.~R.} \bibnamefont{Hamilton}},
  \bibnamefont{and} \bibinfo{author}{\bibfnamefont{R.~G.} \bibnamefont{Clark}},
  \bibinfo{journal}{Phys. Rev. Lett.} \textbf{\bibinfo{volume}{92}},
  \bibinfo{pages}{097901} (\bibinfo{year}{2004}),
  \urlprefix\url{https://link.aps.org/doi/10.1103/PhysRevLett.92.097901}.

\bibitem[{\citenamefont{Collins et~al.}(2002)\citenamefont{Collins, Gisin,
  Linden, Massar, and Popescu}}]{col}
\bibinfo{author}{\bibfnamefont{D.}~\bibnamefont{Collins}},
  \bibinfo{author}{\bibfnamefont{N.}~\bibnamefont{Gisin}},
  \bibinfo{author}{\bibfnamefont{N.}~\bibnamefont{Linden}},
  \bibinfo{author}{\bibfnamefont{S.}~\bibnamefont{Massar}}, \bibnamefont{and}
  \bibinfo{author}{\bibfnamefont{S.}~\bibnamefont{Popescu}},
  \bibinfo{journal}{Phys. Rev. Lett.} \textbf{\bibinfo{volume}{88}},
  \bibinfo{pages}{040404} (\bibinfo{year}{2002}),
  \urlprefix\url{https://link.aps.org/doi/10.1103/PhysRevLett.88.040404}.

\bibitem[{\citenamefont{Kaszlikowski et~al.}(2002)\citenamefont{Kaszlikowski,
  Kwek, Chen, \ifmmode~\dot{Z}\else \.{Z}\fi{}ukowski, and Oh}}]{KKC+02}
\bibinfo{author}{\bibfnamefont{D.}~\bibnamefont{Kaszlikowski}},
  \bibinfo{author}{\bibfnamefont{L.~C.} \bibnamefont{Kwek}},
  \bibinfo{author}{\bibfnamefont{J.-L.} \bibnamefont{Chen}},
  \bibinfo{author}{\bibfnamefont{M.}~\bibnamefont{\ifmmode~\dot{Z}\else
  \.{Z}\fi{}ukowski}}, \bibnamefont{and} \bibinfo{author}{\bibfnamefont{C.~H.}
  \bibnamefont{Oh}}, \bibinfo{journal}{Phys. Rev. A}
  \textbf{\bibinfo{volume}{65}}, \bibinfo{pages}{032118}
  (\bibinfo{year}{2002}),
  \urlprefix\url{https://link.aps.org/doi/10.1103/PhysRevA.65.032118}.

\bibitem[{\citenamefont{Ac\'{\i}n et~al.}(2002)\citenamefont{Ac\'{\i}n, Durt,
  Gisin, and Latorre}}]{acin}
\bibinfo{author}{\bibfnamefont{A.}~\bibnamefont{Ac\'{\i}n}},
  \bibinfo{author}{\bibfnamefont{T.}~\bibnamefont{Durt}},
  \bibinfo{author}{\bibfnamefont{N.}~\bibnamefont{Gisin}}, \bibnamefont{and}
  \bibinfo{author}{\bibfnamefont{J.~I.} \bibnamefont{Latorre}},
  \bibinfo{journal}{Phys. Rev. A} \textbf{\bibinfo{volume}{65}},
  \bibinfo{pages}{052325} (\bibinfo{year}{2002}),
  \urlprefix\url{https://link.aps.org/doi/10.1103/PhysRevA.65.052325}.

\bibitem[{\citenamefont{\ifmmode~\dot{Z}\else \.{Z}\fi{}yczkowski
  et~al.}(1998)\citenamefont{\ifmmode~\dot{Z}\else \.{Z}\fi{}yczkowski,
  Horodecki, Sanpera, and Lewenstein}}]{ZHS+98}
\bibinfo{author}{\bibfnamefont{K.}~\bibnamefont{\ifmmode~\dot{Z}\else
  \.{Z}\fi{}yczkowski}},
  \bibinfo{author}{\bibfnamefont{P.}~\bibnamefont{Horodecki}},
  \bibinfo{author}{\bibfnamefont{A.}~\bibnamefont{Sanpera}}, \bibnamefont{and}
  \bibinfo{author}{\bibfnamefont{M.}~\bibnamefont{Lewenstein}},
  \bibinfo{journal}{Phys. Rev. A} \textbf{\bibinfo{volume}{58}},
  \bibinfo{pages}{883} (\bibinfo{year}{1998}),
  \urlprefix\url{https://link.aps.org/doi/10.1103/PhysRevA.58.883}.

\bibitem[{\citenamefont{Vidal and Werner}(2002)}]{VW02}
\bibinfo{author}{\bibfnamefont{G.}~\bibnamefont{Vidal}} \bibnamefont{and}
  \bibinfo{author}{\bibfnamefont{R.~F.} \bibnamefont{Werner}},
  \bibinfo{journal}{Phys. Rev. A} \textbf{\bibinfo{volume}{65}},
  \bibinfo{pages}{032314} (\bibinfo{year}{2002}),
  \urlprefix\url{https://link.aps.org/doi/10.1103/PhysRevA.65.032314}.

\bibitem[{\citenamefont{Moroder et~al.}(2013)\citenamefont{Moroder, Bancal,
  Liang, Hofmann, and G\"uhne}}]{MBL+13}
\bibinfo{author}{\bibfnamefont{T.}~\bibnamefont{Moroder}},
  \bibinfo{author}{\bibfnamefont{J.-D.} \bibnamefont{Bancal}},
  \bibinfo{author}{\bibfnamefont{Y.-C.} \bibnamefont{Liang}},
  \bibinfo{author}{\bibfnamefont{M.}~\bibnamefont{Hofmann}}, \bibnamefont{and}
  \bibinfo{author}{\bibfnamefont{O.}~\bibnamefont{G\"uhne}},
  \bibinfo{journal}{Phys. Rev. Lett.} \textbf{\bibinfo{volume}{111}},
  \bibinfo{pages}{030501} (\bibinfo{year}{2013}),
  \urlprefix\url{https://link.aps.org/doi/10.1103/PhysRevLett.111.030501}.

\bibitem[{\citenamefont{Pusey}(2013)}]{Pus13}
\bibinfo{author}{\bibfnamefont{M.~F.} \bibnamefont{Pusey}},
  \bibinfo{journal}{Phys. Rev. A} \textbf{\bibinfo{volume}{88}},
  \bibinfo{pages}{032313} (\bibinfo{year}{2013}),
  \urlprefix\url{https://link.aps.org/doi/10.1103/PhysRevA.88.032313}.

\bibitem[{\citenamefont{Eltschka and Siewert}(2013)}]{ES13}
\bibinfo{author}{\bibfnamefont{C.}~\bibnamefont{Eltschka}} \bibnamefont{and}
  \bibinfo{author}{\bibfnamefont{J.}~\bibnamefont{Siewert}},
  \bibinfo{journal}{Phys. Rev. Lett.} \textbf{\bibinfo{volume}{111}},
  \bibinfo{pages}{100503} (\bibinfo{year}{2013}),
  \urlprefix\url{https://link.aps.org/doi/10.1103/PhysRevLett.111.100503}.

\bibitem[{\citenamefont{Horodecki and Horodecki}(1999)}]{HH99}
\bibinfo{author}{\bibfnamefont{M.}~\bibnamefont{Horodecki}} \bibnamefont{and}
  \bibinfo{author}{\bibfnamefont{P.}~\bibnamefont{Horodecki}},
  \bibinfo{journal}{Phys. Rev. A} \textbf{\bibinfo{volume}{59}},
  \bibinfo{pages}{4206} (\bibinfo{year}{1999}),
  \urlprefix\url{https://link.aps.org/doi/10.1103/PhysRevA.59.4206}.

\bibitem[{\citenamefont{Terhal and Vollbrecht}(2000)}]{TV00}
\bibinfo{author}{\bibfnamefont{B.~M.} \bibnamefont{Terhal}} \bibnamefont{and}
  \bibinfo{author}{\bibfnamefont{K.~G.~H.} \bibnamefont{Vollbrecht}},
  \bibinfo{journal}{Phys. Rev. Lett.} \textbf{\bibinfo{volume}{85}},
  \bibinfo{pages}{2625} (\bibinfo{year}{2000}),
  \urlprefix\url{https://link.aps.org/doi/10.1103/PhysRevLett.85.2625}.

\bibitem[{\citenamefont{Rungta and Caves}(2003)}]{RC03}
\bibinfo{author}{\bibfnamefont{P.}~\bibnamefont{Rungta}} \bibnamefont{and}
  \bibinfo{author}{\bibfnamefont{C.~M.} \bibnamefont{Caves}},
  \bibinfo{journal}{Phys. Rev. A} \textbf{\bibinfo{volume}{67}},
  \bibinfo{pages}{012307} (\bibinfo{year}{2003}),
  \urlprefix\url{https://link.aps.org/doi/10.1103/PhysRevA.67.012307}.

\bibitem[{\citenamefont{Popescu}(1994)}]{Pop94}
\bibinfo{author}{\bibfnamefont{S.}~\bibnamefont{Popescu}},
  \bibinfo{journal}{Phys. Rev. Lett.} \textbf{\bibinfo{volume}{72}},
  \bibinfo{pages}{797} (\bibinfo{year}{1994}),
  \urlprefix\url{https://link.aps.org/doi/10.1103/PhysRevLett.72.797}.

\bibitem[{\citenamefont{Eltschka et~al.}(2015)\citenamefont{Eltschka, T\'oth,
  and Siewert}}]{ETS15}
\bibinfo{author}{\bibfnamefont{C.}~\bibnamefont{Eltschka}},
  \bibinfo{author}{\bibfnamefont{G.}~\bibnamefont{T\'oth}}, \bibnamefont{and}
  \bibinfo{author}{\bibfnamefont{J.}~\bibnamefont{Siewert}},
  \bibinfo{journal}{Phys. Rev. A} \textbf{\bibinfo{volume}{91}},
  \bibinfo{pages}{032327} (\bibinfo{year}{2015}),
  \urlprefix\url{https://link.aps.org/doi/10.1103/PhysRevA.91.032327}.

\bibitem[{\citenamefont{Werner}(1989)}]{Wer89}
\bibinfo{author}{\bibfnamefont{R.~F.} \bibnamefont{Werner}},
  \bibinfo{journal}{Phys. Rev. A} \textbf{\bibinfo{volume}{40}},
  \bibinfo{pages}{4277} (\bibinfo{year}{1989}),
  \urlprefix\url{https://link.aps.org/doi/10.1103/PhysRevA.40.4277}.

\bibitem[{\citenamefont{Ac\'{\i}n et~al.}(2005)\citenamefont{Ac\'{\i}n, Gill,
  and Gisin}}]{AGG05}
\bibinfo{author}{\bibfnamefont{A.}~\bibnamefont{Ac\'{\i}n}},
  \bibinfo{author}{\bibfnamefont{R.}~\bibnamefont{Gill}}, \bibnamefont{and}
  \bibinfo{author}{\bibfnamefont{N.}~\bibnamefont{Gisin}},
  \bibinfo{journal}{Phys. Rev. Lett.} \textbf{\bibinfo{volume}{95}},
  \bibinfo{pages}{210402} (\bibinfo{year}{2005}),
  \urlprefix\url{https://link.aps.org/doi/10.1103/PhysRevLett.95.210402}.

\bibitem[{\citenamefont{Brunner et~al.}(2005)\citenamefont{Brunner, Gisin, and
  Scarani}}]{BGS05}
\bibinfo{author}{\bibfnamefont{N.}~\bibnamefont{Brunner}},
  \bibinfo{author}{\bibfnamefont{N.}~\bibnamefont{Gisin}}, \bibnamefont{and}
  \bibinfo{author}{\bibfnamefont{V.}~\bibnamefont{Scarani}},
  \bibinfo{journal}{New Journal of Physics} \textbf{\bibinfo{volume}{7}},
  \bibinfo{pages}{88} (\bibinfo{year}{2005}),
  \urlprefix\url{http://stacks.iop.org/1367-2630/7/i=1/a=088}.

\bibitem[{\citenamefont{Zohren and Gill}(2008)}]{ZG08}
\bibinfo{author}{\bibfnamefont{S.}~\bibnamefont{Zohren}} \bibnamefont{and}
  \bibinfo{author}{\bibfnamefont{R.~D.} \bibnamefont{Gill}},
  \bibinfo{journal}{Phys. Rev. Lett.} \textbf{\bibinfo{volume}{100}},
  \bibinfo{pages}{120406} (\bibinfo{year}{2008}),
  \urlprefix\url{https://link.aps.org/doi/10.1103/PhysRevLett.100.120406}.

\bibitem[{\citenamefont{Junge and Palazuelos}(2011)}]{JP11}
\bibinfo{author}{\bibfnamefont{M.}~\bibnamefont{Junge}} \bibnamefont{and}
  \bibinfo{author}{\bibfnamefont{C.}~\bibnamefont{Palazuelos}},
  \bibinfo{journal}{Communications in Mathematical Physics}
  \textbf{\bibinfo{volume}{306}}, \bibinfo{pages}{695} (\bibinfo{year}{2011}),
  ISSN \bibinfo{issn}{1432-0916},
  \urlprefix\url{https://doi.org/10.1007/s00220-011-1296-8}.

\bibitem[{\citenamefont{Bernhard et~al.}(2014)\citenamefont{Bernhard, Bessire,
  Montina, Pfaffhauser, Stefanov, and Wolf}}]{BBM+14}
\bibinfo{author}{\bibfnamefont{C.}~\bibnamefont{Bernhard}},
  \bibinfo{author}{\bibfnamefont{B.}~\bibnamefont{Bessire}},
  \bibinfo{author}{\bibfnamefont{A.}~\bibnamefont{Montina}},
  \bibinfo{author}{\bibfnamefont{M.}~\bibnamefont{Pfaffhauser}},
  \bibinfo{author}{\bibfnamefont{A.}~\bibnamefont{Stefanov}}, \bibnamefont{and}
  \bibinfo{author}{\bibfnamefont{S.}~\bibnamefont{Wolf}},
  \bibinfo{journal}{Journal of Physics A: Mathematical and Theoretical}
  \textbf{\bibinfo{volume}{47}}, \bibinfo{pages}{424013}
  (\bibinfo{year}{2014}),
  \urlprefix\url{http://stacks.iop.org/1751-8121/47/i=42/a=424013}.

\bibitem[{\citenamefont{Das et~al.}(2017)\citenamefont{Das, Datta, Goswami,
  Majumdar, and Home}}]{DDG+17}
\bibinfo{author}{\bibfnamefont{D.}~\bibnamefont{Das}},
  \bibinfo{author}{\bibfnamefont{S.}~\bibnamefont{Datta}},
  \bibinfo{author}{\bibfnamefont{S.}~\bibnamefont{Goswami}},
  \bibinfo{author}{\bibfnamefont{A.}~\bibnamefont{Majumdar}}, \bibnamefont{and}
  \bibinfo{author}{\bibfnamefont{D.}~\bibnamefont{Home}},
  \bibinfo{journal}{Physics Letters A} \textbf{\bibinfo{volume}{381}},
  \bibinfo{pages}{3396 } (\bibinfo{year}{2017}), ISSN
  \bibinfo{issn}{0375-9601},
  \urlprefix\url{http://www.sciencedirect.com/science/article/pii/S0375960117308101}.

\bibitem[{\citenamefont{Nielsen and Chuang}(2000)}]{NC00}
\bibinfo{author}{\bibfnamefont{M.~A.} \bibnamefont{Nielsen}} \bibnamefont{and}
  \bibinfo{author}{\bibfnamefont{I.~L.} \bibnamefont{Chuang}},
  \emph{\bibinfo{title}{Quantum Computation and Quantum Information}}
  (\bibinfo{publisher}{Cambridge University Press, Cambridge, England},
  \bibinfo{year}{2000}).

\bibitem[{\citenamefont{Scarani et~al.}(2006)\citenamefont{Scarani, Gisin,
  Brunner, Masanes, Pino, and Ac\'{\i}n}}]{SGB+06}
\bibinfo{author}{\bibfnamefont{V.}~\bibnamefont{Scarani}},
  \bibinfo{author}{\bibfnamefont{N.}~\bibnamefont{Gisin}},
  \bibinfo{author}{\bibfnamefont{N.}~\bibnamefont{Brunner}},
  \bibinfo{author}{\bibfnamefont{L.}~\bibnamefont{Masanes}},
  \bibinfo{author}{\bibfnamefont{S.}~\bibnamefont{Pino}}, \bibnamefont{and}
  \bibinfo{author}{\bibfnamefont{A.}~\bibnamefont{Ac\'{\i}n}},
  \bibinfo{journal}{Phys. Rev. A} \textbf{\bibinfo{volume}{74}},
  \bibinfo{pages}{042339} (\bibinfo{year}{2006}),
  \urlprefix\url{https://link.aps.org/doi/10.1103/PhysRevA.74.042339}.

\bibitem[{\citenamefont{Peres}(1996)}]{Per96}
\bibinfo{author}{\bibfnamefont{A.}~\bibnamefont{Peres}},
  \bibinfo{journal}{Phys. Rev. Lett.} \textbf{\bibinfo{volume}{77}},
  \bibinfo{pages}{1413} (\bibinfo{year}{1996}),
  \urlprefix\url{https://link.aps.org/doi/10.1103/PhysRevLett.77.1413}.

\bibitem[{\citenamefont{Sent\'{\i}s et~al.}(2016)\citenamefont{Sent\'{\i}s,
  Eltschka, G\"uhne, Huber, and Siewert}}]{SEG+16}
\bibinfo{author}{\bibfnamefont{G.}~\bibnamefont{Sent\'{\i}s}},
  \bibinfo{author}{\bibfnamefont{C.}~\bibnamefont{Eltschka}},
  \bibinfo{author}{\bibfnamefont{O.}~\bibnamefont{G\"uhne}},
  \bibinfo{author}{\bibfnamefont{M.}~\bibnamefont{Huber}}, \bibnamefont{and}
  \bibinfo{author}{\bibfnamefont{J.}~\bibnamefont{Siewert}},
  \bibinfo{journal}{Phys. Rev. Lett.} \textbf{\bibinfo{volume}{117}},
  \bibinfo{pages}{190502} (\bibinfo{year}{2016}),
  \urlprefix\url{https://link.aps.org/doi/10.1103/PhysRevLett.117.190502}.

\bibitem[{\citenamefont{Quintino et~al.}(2015)\citenamefont{Quintino,
  V\'ertesi, Cavalcanti, Augusiak, Demianowicz, Ac\'{\i}n, and
  Brunner}}]{QVC+15}
\bibinfo{author}{\bibfnamefont{M.~T.} \bibnamefont{Quintino}},
  \bibinfo{author}{\bibfnamefont{T.}~\bibnamefont{V\'ertesi}},
  \bibinfo{author}{\bibfnamefont{D.}~\bibnamefont{Cavalcanti}},
  \bibinfo{author}{\bibfnamefont{R.}~\bibnamefont{Augusiak}},
  \bibinfo{author}{\bibfnamefont{M.}~\bibnamefont{Demianowicz}},
  \bibinfo{author}{\bibfnamefont{A.}~\bibnamefont{Ac\'{\i}n}},
  \bibnamefont{and} \bibinfo{author}{\bibfnamefont{N.}~\bibnamefont{Brunner}},
  \bibinfo{journal}{Phys. Rev. A} \textbf{\bibinfo{volume}{92}},
  \bibinfo{pages}{032107} (\bibinfo{year}{2015}),
  \urlprefix\url{https://link.aps.org/doi/10.1103/PhysRevA.92.032107}.

\bibitem[{\citenamefont{Shen et~al.}(2015)\citenamefont{Shen, Li, and
  Duan}}]{SLD15}
\bibinfo{author}{\bibfnamefont{S.-Q.} \bibnamefont{Shen}},
  \bibinfo{author}{\bibfnamefont{M.}~\bibnamefont{Li}}, \bibnamefont{and}
  \bibinfo{author}{\bibfnamefont{X.-F.} \bibnamefont{Duan}},
  \bibinfo{journal}{Phys. Rev. A} \textbf{\bibinfo{volume}{91}},
  \bibinfo{pages}{012326} (\bibinfo{year}{2015}),
  \urlprefix\url{https://link.aps.org/doi/10.1103/PhysRevA.91.012326}.

\bibitem[{\citenamefont{Wu et~al.}(2014)\citenamefont{Wu, Ma, Chen, and
  Yu}}]{WMC+14}
\bibinfo{author}{\bibfnamefont{S.}~\bibnamefont{Wu}},
  \bibinfo{author}{\bibfnamefont{Z.}~\bibnamefont{Ma}},
  \bibinfo{author}{\bibfnamefont{Z.}~\bibnamefont{Chen}}, \bibnamefont{and}
  \bibinfo{author}{\bibfnamefont{S.}~\bibnamefont{Yu}}, \bibinfo{journal}{Sci.
  Rep} \textbf{\bibinfo{volume}{4}}, \bibinfo{pages}{4036}
  (\bibinfo{year}{2014}).

\bibitem[{\citenamefont{Guo and Wu}(2014)}]{GW14}
\bibinfo{author}{\bibfnamefont{Y.}~\bibnamefont{Guo}} \bibnamefont{and}
  \bibinfo{author}{\bibfnamefont{S.}~\bibnamefont{Wu}}, \bibinfo{journal}{Sci.
  Rep} \textbf{\bibinfo{volume}{4}}, \bibinfo{pages}{7179}
  (\bibinfo{year}{2014}).

\bibitem[{\citenamefont{Bennett et~al.}(1996)\citenamefont{Bennett, DiVincenzo,
  Smolin, and Wootters}}]{BDS+96}
\bibinfo{author}{\bibfnamefont{C.~H.} \bibnamefont{Bennett}},
  \bibinfo{author}{\bibfnamefont{D.~P.} \bibnamefont{DiVincenzo}},
  \bibinfo{author}{\bibfnamefont{J.~A.} \bibnamefont{Smolin}},
  \bibnamefont{and} \bibinfo{author}{\bibfnamefont{W.~K.}
  \bibnamefont{Wootters}}, \bibinfo{journal}{Phys. Rev. A}
  \textbf{\bibinfo{volume}{54}}, \bibinfo{pages}{3824} (\bibinfo{year}{1996}),
  \urlprefix\url{https://link.aps.org/doi/10.1103/PhysRevA.54.3824}.

\bibitem[{\citenamefont{Horodecki et~al.}(2009)\citenamefont{Horodecki,
  Horodecki, Horodecki, and Horodecki}}]{HHH+09}
\bibinfo{author}{\bibfnamefont{R.}~\bibnamefont{Horodecki}},
  \bibinfo{author}{\bibfnamefont{P.}~\bibnamefont{Horodecki}},
  \bibinfo{author}{\bibfnamefont{M.}~\bibnamefont{Horodecki}},
  \bibnamefont{and}
  \bibinfo{author}{\bibfnamefont{K.}~\bibnamefont{Horodecki}},
  \bibinfo{journal}{Rev. Mod. Phys.} \textbf{\bibinfo{volume}{81}},
  \bibinfo{pages}{865} (\bibinfo{year}{2009}),
  \urlprefix\url{https://link.aps.org/doi/10.1103/RevModPhys.81.865}.

\bibitem[{\citenamefont{Eisert and Plenio}(1999)}]{EP99}
\bibinfo{author}{\bibfnamefont{J.}~\bibnamefont{Eisert}} \bibnamefont{and}
  \bibinfo{author}{\bibfnamefont{M.~B.} \bibnamefont{Plenio}},
  \bibinfo{journal}{Journal of Modern Optics} \textbf{\bibinfo{volume}{46}},
  \bibinfo{pages}{145} (\bibinfo{year}{1999}).

\bibitem[{\citenamefont{\ifmmode~\dot{Z}\else
  \.{Z}\fi{}yczkowski}(1999)}]{Zyc99}
\bibinfo{author}{\bibfnamefont{K.}~\bibnamefont{\ifmmode~\dot{Z}\else
  \.{Z}\fi{}yczkowski}}, \bibinfo{journal}{Phys. Rev. A}
  \textbf{\bibinfo{volume}{60}}, \bibinfo{pages}{3496} (\bibinfo{year}{1999}),
  \urlprefix\url{https://link.aps.org/doi/10.1103/PhysRevA.60.3496}.

\bibitem[{\citenamefont{Miranowicz and Grudka}(2004{\natexlab{a}})}]{MG04}
\bibinfo{author}{\bibfnamefont{A.}~\bibnamefont{Miranowicz}} \bibnamefont{and}
  \bibinfo{author}{\bibfnamefont{A.}~\bibnamefont{Grudka}},
  \bibinfo{journal}{Phys. Rev. A} \textbf{\bibinfo{volume}{70}},
  \bibinfo{pages}{032326} (\bibinfo{year}{2004}{\natexlab{a}}),
  \urlprefix\url{https://link.aps.org/doi/10.1103/PhysRevA.70.032326}.

\bibitem[{\citenamefont{Miranowicz and Grudka}(2004{\natexlab{b}})}]{MG'04}
\bibinfo{author}{\bibfnamefont{A.}~\bibnamefont{Miranowicz}} \bibnamefont{and}
  \bibinfo{author}{\bibfnamefont{A.}~\bibnamefont{Grudka}},
  \bibinfo{journal}{J. Opt. B: Quantum Semiclass. Opt.}
  \textbf{\bibinfo{volume}{6}}, \bibinfo{pages}{542}
  (\bibinfo{year}{2004}{\natexlab{b}}),
  \urlprefix\url{http://stacks.iop.org/1464-4266/6/i=12/a=009}.

\bibitem[{\citenamefont{Verstraete et~al.}(2001)\citenamefont{Verstraete,
  Audenaert, Dehaene, and Moor}}]{VAD01}
\bibinfo{author}{\bibfnamefont{F.}~\bibnamefont{Verstraete}},
  \bibinfo{author}{\bibfnamefont{K.}~\bibnamefont{Audenaert}},
  \bibinfo{author}{\bibfnamefont{J.}~\bibnamefont{Dehaene}}, \bibnamefont{and}
  \bibinfo{author}{\bibfnamefont{B.~D.} \bibnamefont{Moor}},
  \bibinfo{journal}{Journal of Physics A: Mathematical and General}
  \textbf{\bibinfo{volume}{34}}, \bibinfo{pages}{10327} (\bibinfo{year}{2001}),
  \urlprefix\url{http://stacks.iop.org/0305-4470/34/i=47/a=329}.

\bibitem[{\citenamefont{Chattopadhyay and Sarkar}(2008)}]{CS08}
\bibinfo{author}{\bibfnamefont{I.}~\bibnamefont{Chattopadhyay}}
  \bibnamefont{and} \bibinfo{author}{\bibfnamefont{D.}~\bibnamefont{Sarkar}},
  \bibinfo{journal}{Quantum Information Processing}
  \textbf{\bibinfo{volume}{7}}, \bibinfo{pages}{243} (\bibinfo{year}{2008}),
  ISSN \bibinfo{issn}{1573-1332},
  \urlprefix\url{https://doi.org/10.1007/s11128-008-0085-6}.

\bibitem[{\citenamefont{Erol}(2015)}]{Ero15}
\bibinfo{author}{\bibfnamefont{V.}~\bibnamefont{Erol}}, \bibinfo{journal}{AIP
  Conference Proceedings} \textbf{\bibinfo{volume}{1653}},
  \bibinfo{pages}{020037} (\bibinfo{year}{2015}).

\end{thebibliography}

\appendix

\section{Derivation of Eq. (\ref{d3isoNS}) for the sum of four 
PCCs for the two-qutrit isotropic mixed states}  \label{AIII}
For  $A_1=B_1=\sum_j a_j \ketbra{a_j}{a_j}$ in which the basis $\{\ket{a_j}\}$ is the computational basis
and the eigenvalues $a_j$ are given by $a_0=+1$, $a_1=0$ and $a_2=-1$, the relevant single and joint
expectation values of the two-qutrit isotropic  states given by Eq. (\ref{nqudiso}) with $d=3$ are given by
\begin{align}
 \braket{A_1}&=\braket{B_1}=0,  \nonumber \\
 \braket{A^2_1}&=\braket{B^2_1}=\frac{2}{3},  \nonumber \\ 
 \braket{A_1B_1}&=\frac{-1+9p}{12}. \nonumber
\end{align}
From the above expressions, it can be checked that the PCC in this case takes the value
\be
C_{A_1B_1}=\frac{-1+9p}{8}. \label{AIII1}
\ee

For  $A_2=B_2=\sum_j b_j \ketbra{b_j}{b_j}$, where the basis $\{\ket{b_j}\}$ is given in Eq. (\ref{csmub3}) 
and the eigenvalues $b_j$ are given by $b_0=0$, $b_1=\pm 1$ and $b_2= \mp 1$, the relevant  single and joint
expectation values are given by
\begin{align}
 \braket{A_2}&=\braket{B_2}=0, \nonumber \\
 \braket{A^2_2}&=\braket{B^2_2}=\frac{2}{3},  \nonumber \\ 
 \braket{A_2B_2}&=\frac{1-9p}{12}. \nonumber
\end{align}
From the above expressions, it can be checked that the PCC  in this case is given by
\be
C_{A_2B_2}=\frac{1-9p}{8}. \label{AIII2}
\ee

For  $A_3=B_3=\sum_j e_j \ketbra{e_j}{e_j}$, where the basis $\{\ket{e_j}\}$ is given in Eq. (\ref{csmub3}) and
 the eigenvalues $e_j$ are given by $e_0=+1$, $e_1=0$ and $e_2=-1$, the relevant  single and joint
expectation values are given by
\begin{align}
 \braket{A_3}&=\braket{B_3}=0, \nonumber \\
 \braket{A^2_3}&=\braket{B^2_3}=\frac{2}{3}, \nonumber \\ 
 \braket{A_3B_3}&=\frac{1-9p}{12}. \nonumber
\end{align}
From the above expressions, it can be checked that the PCC in this case takes the value
\be
C_{A_3B_3}=\frac{1-9p}{8}. \label{AIII3}
\ee

For  $A_4=B_4=\sum_j g_j \ketbra{g_j}{g_j}$, where the basis $\{\ket{g_j}\}$ is given in Eq. (\ref{csmub3}) and
the eigenvalues $g_j$ are given by $g_0=+1$, $g_1=0$ and $g_2=-1$, the relevant single and joint
expectation values are given by
\begin{align}
 \braket{A_4}&=\braket{B_4}=0, \nonumber \\
 \braket{A^2_4}&=\braket{B^2_4}=\frac{2}{3}, \nonumber \\ 
 \braket{A_4B_4}&=\frac{1-9p}{12}. \nonumber
\end{align}
From the above expressions, it can be checked that the PCC in this case is given by
\be
C_{A_4B_4}=\frac{1-9p}{8}. \label{AIII4}
\ee
Then  Eq. (\ref{d3isoNS}) follows from Eqs. (\ref{AIII1})-(\ref{AIII4}).

\section{Derivation of Eq. (\ref{PCCcc}) for the sum of two
PCCs for the coloured-noise two-qutrit maximally entangled state-A} \label{AIV}
For  $A_1=B_1=\sum_j a_j \ketbra{a_j}{a_j}$ in which the basis $\{\ket{a_j}\}$ is the computational basis and the eigenvalues $a_j$ are given by $a_0=+1$, $a_1=0$ and $a_2=-1$, 
the relevant single and joint
expectation values for the coloured-noise two-qutrit maximally entangled state given by Eq. (\ref{cndme}) with $d=3$ are given by
\begin{align}
 \braket{A_1}&=\braket{B_1}=0, \nonumber \\
 \braket{A^2_1}&=\braket{B^2_1}=2/3,  \nonumber \\ 
 \braket{A_1B_1}&=2/3. \nonumber 
\end{align}
From the above expressions, it can be checked that the PCC in this case takes the value
\be
C_{A_1B_1}=1. \label{AIV1}
\ee

For  $A_2=B_2=\sum_j b_j \ketbra{b_j}{b_j}$ in which the basis $\{\ket{b_j}\}$ is the generalized $\sigma_y$ basis and
 the eigenvalues $b_j$ are given by $b_0=+1$, $b_1=0$ and $b_2=-1$, the relevant  single and joint
expectations are given by
\begin{align}
 \braket{A_2}&=\braket{B_2}=0,  \nonumber \\
 \braket{A^2_2}&=\braket{B^2_2}=\frac{2}{3},  \nonumber \\ 
 \braket{A_2B_2}&=\frac{-2p}{3}. \nonumber
\end{align}
From the above expressions, it can be checked that the PCC in this case is given by
\be
C_{A_2B_2}=-p. \label{AIV2}
\ee
Then Eq. (\ref{PCCcc})  follows from Eqs. (\ref{AIV1}) and (\ref{AIV2}).

\section{Derivation of Eq. (\ref{PCCcc1}) for the sum of four 
PCCs for the coloured-noise two-qutrit maximally entangled state-B}  \label{ACNB}
For  $A_1=B_1=\sum_j a_j \ketbra{a_j}{a_j}$ in which the basis $\{\ket{a_j}\}$ is the computational basis and the eigenvalues $a_j$ are given by $a_0=+1$, $a_1=0$ and $a_2=-1$, the relevant single and joint
expectation values of the coloured-noise two-qutrit maximally entangled state given by Eq. (\ref{cndme1}) with $d=3$ are given by
\begin{align}
 \braket{A_1}&=\braket{B_1}=0,  \nonumber \\
 \braket{A^2_1}&=\braket{B^2_1}=\frac{2}{3},  \nonumber \\ 
 \braket{A_1B_1}&=\frac{-1+3p}{3}. \nonumber
\end{align}
From the above expressions, it can be checked that the PCC in this case takes the value
\be
C_{A_1B_1}=\frac{-1+3p}{2}. \label{ACNB1}
\ee

For  $A_2=B_2=\sum_j b_j \ketbra{b_j}{b_j}$, where the basis $\{\ket{b_j}\}$ is given in Eq. (\ref{csmub3}) and 
the eigenvalues $b_j$ are given by $b_0=0$, $b_1=\pm 1$ and $b_2= \mp 1$, the relevant  single and joint
expectation values are given by
\begin{align}
 \braket{A_2}&=\braket{B_2}=0, \nonumber \\
 \braket{A^2_2}&=\braket{B^2_2}=\frac{2}{3},  \nonumber \\ 
 \braket{A_2B_2}&=\frac{-2p}{3}. \nonumber
\end{align}
From the above expressions, it can be checked that the PCC  in this case is given by
\be
C_{A_2B_2}=-p. \label{ACNB2}
\ee

For  $A_3=B_3=\sum_j e_j \ketbra{e_j}{e_j}$, where the basis $\{\ket{e_j}\}$ is given in Eq. (\ref{csmub3})
and the eigenvalues $e_j$ are given by $e_0=+1$, $e_1=0$ and $e_2=-1$, the relevant  single and joint
expectation values are given by
\begin{align}
 \braket{A_3}&=\braket{B_3}=0, \nonumber \\
 \braket{A^2_3}&=\braket{B^2_3}=\frac{2}{3}, \nonumber \\ 
 \braket{A_3B_3}&=\frac{-2p}{3}. \nonumber
\end{align}
From the above expressions, it can be checked that the PCC in this case takes the value
\be
C_{A_3B_3}=-p. \label{ACNB3}
\ee

For  $A_4=B_4=\sum_j g_j \ketbra{g_j}{g_j}$, where the basis $\{\ket{g_j}\}$ is given in Eq. (\ref{csmub3})
and the eigenvalues $g_j$ are given by $g_0=+1$, $g_1=0$ and $g_2=-1$, the relevant single and joint
expectation values are given by
\begin{align}
 \braket{A_4}&=\braket{B_4}=0, \nonumber \\
 \braket{A^2_4}&=\braket{B^2_4}=\frac{2}{3}, \nonumber \\ 
 \braket{A_4B_4}&=\frac{-2p}{3}. \nonumber
\end{align}
From the above expressions, it can be checked that the PCC in this case is given by
\be
C_{A_4B_4}=-p. \label{ACNB4}
\ee
Then  Eq. (\ref{PCCcc1}) follows from Eqs. (\ref{ACNB1})-(\ref{ACNB4}).

\section{Derivation of Eq. (\ref{sPw})  for the sum of 
PCCs for the two-qutrit Werner states}  \label{AV}
For  $A_1=B_1=\sum_j a_j \ketbra{a_j}{a_j}$ in which the basis $\{\ket{a_j}\}$ is the computational basis
and the eigenvalues $a_j$ are given by $a_0=+1$, $a_1=0$ and $a_2=-1$, the relevant single and joint
expectations of the two-qutrit Werner  states given by Eq. (\ref{wernerd}) with $d=3$ are given by
\begin{align}
 \braket{A_1}&=\braket{B_1}=0, \nonumber \\
 \braket{A^2_1}&=\braket{B^2_1}=\frac{2}{3}, \nonumber \\ 
 \braket{A_1B_1}&=-\frac{p}{3}. \nonumber
\end{align}
From the above expressions, it can be checked that the PCC in this case takes the value
\be
C_{A_1B_1}=\frac{-p}{2}. \label{AV1}
\ee

For  $A_2=B_2=\sum_j b_j \ketbra{b_j}{b_j}$, where the basis $\{\ket{b_j}\}$ is given in Eq. (\ref{csmub3})
and the eigenvalues $b_j$ are given by $b_0=0$, $b_1=\pm1$ and $b_2=\mp1$, the relevant single and joint
expectations are given by
\begin{align}
 \braket{A_2}&=\braket{B_2}=0, \nonumber \\
 \braket{A^2_2}&=\braket{B^2_2}=\frac{2}{3}, \nonumber \\ 
 \braket{A_2B_2}&=-\frac{p}{3}. \nonumber
\end{align}
From the above expressions, it can be checked that the PCC in this case is given by
\be
C_{A_1B_1}=\frac{-p}{2}. \label{AV2}
\ee

For  $A_3=B_3=\sum_j e_j \ketbra{e_j}{e_j}$, where the basis $\{\ket{e_j}\}$ is given in Eq. (\ref{csmub3})
and the eigenvalues $e_j$ are given by $e_0=+1$, $e_1=0$ and $e_2=-1$, the relevant single and joint
expectations are given by
\begin{align}
 \braket{A_3}&=\braket{B_3}=0, \nonumber \\
 \braket{A^2_3}&=\braket{B^2_3}=\frac{2}{3}, \nonumber \\ 
 \braket{A_3B_3}&=-\frac{p}{3}. \nonumber
\end{align}
From the above expressions, it can be checked that the PCC in this case takes the value
\be
C_{A_3B_3}=\frac{-p}{2}. \label{AV3}
\ee

For  $A_4=B_4=\sum_j g_j \ketbra{g_j}{g_j}$, where the basis $\{\ket{g_j}\}$ is given in Eq. (\ref{csmub3})
and the eigenvalues $g_j$ are given by $g_0=+1$, $g_1=0$ and $g_2=-1$, the relevant single and joint
expectations are given by
\begin{align}
 \braket{A_4}&=\braket{B_4}=0, \nonumber \\
 \braket{A^2_4}&=\braket{B^2_4}=\frac{2}{3}, \nonumber \\ 
 \braket{A_4B_4}&=-\frac{p}{3}. \nonumber
\end{align}
From the above expressions, it can be checked that the PCC in this case is given by
\be
C_{A_4B_4}=\frac{-p}{2}. \label{AV4}
\ee
Then Eq. (\ref{sPw})  follows from Eqs. (\ref{AV1})-(\ref{AV4}).

\section{$d+1$ mutually unbiased bases which can be used for certifying entanglement of $d=3$ and $5$ Werner states} \label{MUBsd35}
To obtain the expression for the sum of $4$ PCCs given in Eq. (\ref{sPw}) for the two-qutrit Werner states,
one can also use the following $4$ mutually unbiased bases:
\begin{align} \label{csmubd=4}
\{\ket{a_j}\}=&\{\ket{0}, \ket{1}, \ket{2} \} \nonumber \\
\{\ket{b_j}\}=&\{(\ket{0}+\omega \ket{1}+\omega^2\ket{2})/\sqrt{3}, \nonumber \\
                &(\ket{0}+ \ket{1}+\ket{2})/\sqrt{3} , \nonumber \\
                 &(\ket{0}+ \omega^2 \ket{1}+ \omega\ket{2})/\sqrt{3}\}, \nonumber \\
\{\ket{e_j}\}=&\{(\ket{0}+\omega \ket{1}+\omega \ket{2})/\sqrt{3}, \nonumber \\
                & (\ket{0}+ \ket{1}+\omega^2 \ket{2})/\sqrt{3}, \nonumber \\
                 &(\ket{0}+\omega^2  \ket{1}+ \ket{2})/\sqrt{3}\}, \nonumber \\
\{\ket{g_j}\}=&\{(\ket{0}+\omega^2 \ket{1}+\omega^2 \ket{2})/\sqrt{3}, \nonumber \\
                & (\ket{0}+ \ket{1}+\omega \ket{2})/\sqrt{3}, \nonumber \\
                 &(\ket{0}+\omega  \ket{1}+ \ket{2})/\sqrt{3}\},               
\end{align}
where $\omega=e^{2i\pi/3}$. 

\vspace{2cm}
The expression obtained for the sum of $6$ PCCs in Eq. (\ref{sPw}) for the  Werner states in $d=5$,
can also be obtained by using the following $6$ mutually unbiased bases:
\begin{widetext}
\begin{align}\label{csmubd=5}
 \{\ket{a_j}\}=&\{\ket{0}, \ket{1}, \ket{2}, \ket{3}, \ket{4}\} \nonumber \\
 \{\ket{b_j}\}=&\{(\ket{0}+ \omega^3 \ket{1}+  \omega \ket{2}+  \omega^4 \ket{3}+ \omega^2 \ket{4})/\sqrt{5},  \nonumber \\
                 &(\ket{0}+ \omega^4 \ket{1}+  \omega^3 \ket{2}+  \omega^2 \ket{3}+ \omega \ket{4})/\sqrt{5},  \nonumber \\
                 &(\ket{0}+  \ket{1}+\ket{2}+\ket{3}+\ket{4})/\sqrt{5}, \nonumber \\
                & (\ket{0}+ \omega \ket{1}+ \omega^2 \ket{2}+ \omega^3 \ket{3}+ \omega^4 \ket{4})/\sqrt{5}, \nonumber \\
                 &(\ket{0}+ \omega^2 \ket{1}+  \omega^4 \ket{2}+  \omega \ket{3}+ \omega^3 \ket{4})/\sqrt{5}\} \nonumber \\
  \{\ket{e_j}\}=&\{(\ket{0}+ \omega^3 \ket{1}+ \omega^3 \ket{2}+ \ket{3}+ \omega^4 \ket{4})/\sqrt{5}, \nonumber \\
                & (\ket{0}+ \omega^4 \ket{1}+  \ket{2}+ \omega^3   \ket{3}  + \omega^3 \ket{4})/\sqrt{5}, \nonumber \\
                 &(\ket{0}+\ket{1}+  \omega^2 \ket{2}+  \omega  \ket{3}+ \omega^2 \ket{4})/\sqrt{5}, \nonumber \\
                 &(\ket{0}+ \omega  \ket{1}+  \omega^4 \ket{2}+  \omega^4  \ket{3}+ \omega \ket{4})/\sqrt{5},  \nonumber \\
                 &(\ket{0}+ \omega^2  \ket{1}+  \omega \ket{2}+  \omega^2  \ket{3}+  \ket{4})/\sqrt{5}\} \nonumber \\ 
   \{\ket{g_j}\}=&\{(\ket{0}+ \omega^4 \ket{1}+  \omega^2 \ket{2}+  \omega^4 \ket{3}+  \ket{4})/\sqrt{5},  \nonumber \\
                 &( \ket{0}+  \ket{1}+  \omega^4 \ket{2}+  \omega^2 \ket{3}+ \omega^4 \ket{4})/\sqrt{5},  \nonumber \\
                 &(\ket{0}+  \omega \ket{1}+ \omega \ket{2}+\ket{3}+ \omega^3 \ket{4})/\sqrt{5}, \nonumber \\
                 &(\ket{0}+ \omega^2 \ket{1}+ \omega^3 \ket{2}+ \omega^3 \ket{3}+ \omega^2 \ket{4})/\sqrt{5}, \nonumber \\
                 &(\ket{0}+ \omega^4 \ket{1}+  \ket{2}+  \omega \ket{3}+ \omega \ket{4})/\sqrt{5}\} \nonumber \\
  \{\ket{k_j}\}=&\{(\ket{0}+  \ket{1}+ \omega \ket{2}+\omega^4 \ket{3}+\omega \ket{4})/\sqrt{5}, \nonumber \\
                & (\ket{0}+ \omega  \ket{1}+ \omega^4 \ket{2}+ \omega \ket{3}+  \ket{4})/\sqrt{5}, \nonumber \\
                 &( \ket{0}+ \omega^2  \ket{1}+   \ket{2}+  \omega^4 \ket{3}+ \omega^4 \ket{4})/\sqrt{5}, \nonumber \\
                 &(\ket{0}+ \omega^4 \ket{1}+  \omega^2 \ket{2}+  \omega^2   \ket{3}+ \omega^4 \ket{4})/\sqrt{5},  \nonumber \\
                 &(\ket{0}+ \omega^4  \ket{1}+  \omega^4  \ket{2}+  \ket{3}+  \omega^2  \ket{4})/\sqrt{5}\} \nonumber \\
 \{\ket{l_j}\}=&\{(  \ket{0}+ \omega  \ket{1}+  \ket{2}+   \omega^2  \ket{3}+ \omega^2 \ket{4})/\sqrt{5},  \nonumber \\
                 &( \ket{0}+ \omega^2 \ket{1}+  \omega^2 \ket{2}+   \ket{3}+  \omega \ket{4})/\sqrt{5}, \nonumber \\
                 &(\ket{0}+ \omega^3 \ket{1}+\omega^4 \ket{2}+ \omega^4 \ket{3}+  \ket{4})/\sqrt{5}, \nonumber \\
                & (\ket{0}+ \omega^4 \ket{1}+   \omega \ket{2}+ \omega \ket{3}+\omega^4 \ket{4})/\sqrt{5}, \nonumber \\
                 &( \ket{0}+  \ket{1}+  \omega^3   \ket{2}+ \omega^4 \ket{3}+ \omega^3 \ket{4})/\sqrt{5}\},
\end{align}
\end{widetext}
where $\omega=2i\pi/5$.  

\newpage
\section{$d+1$ noncommuting bases used for calculating the sum of $d+1$ PCCs in the case of $d=4$ and $5$ isotropic and Werner states} \label{NCB45}
For calculating the sum of $5$ PCCs in the case of $d=4$ isotropic states and Werner states,
we consider the following choice of $5$ mutually unbiased bases:
\begin{align}\label{csmub4}
 \{\ket{a_j}\}=&\{\ket{0}, \ket{1}, \ket{2}, \ket{3} \} \nonumber \\
 \{\ket{b_j}\}=&\{(\ket{0}+ \ket{1}+\ket{2}+\ket{3})/2, \nonumber \\
                & (\ket{0}+ \ket{1}-\ket{2}-\ket{3})/2, \nonumber \\
                 &(\ket{0}- \ket{1}-\ket{2}+\ket{3})/2, \nonumber \\
                 &(\ket{0}- \ket{1}+\ket{2}-\ket{3})/2\}, \nonumber \\
  \{\ket{e_j}\}=&\{(\ket{0}+ \ket{1}+ i \ket{2}-i \ket{3})/2, \nonumber \\
                &(\ket{0}- \ket{1}+ i \ket{2}+ i \ket{3})/2, \nonumber \\
                 &(\ket{0}- \ket{1} -i \ket{2}-i \ket{3})/2, \nonumber \\
                 &(\ket{0}+ \ket{1}-i \ket{2}+i\ket{3})/2\}, \nonumber \\  
  \{\ket{g_j}\}=&\{(\ket{0}-i \ket{1}- \ket{2}-i \ket{3})/2, \nonumber \\
                & (\ket{0}+i \ket{1}+  \ket{2}- i \ket{3})/2, \nonumber \\
                 &(\ket{0}-i \ket{1} + \ket{2}+i \ket{3})/2, \nonumber \\
                 &(\ket{0}+ i \ket{1}- \ket{2}+i\ket{3})/2\} \nonumber \\
 \{\ket{k_j}\}=&\{(\ket{0}+i \ket{1} -i  \ket{2} + \ket{3})/2, \nonumber \\
                 &(\ket{0}-i \ket{1} + i \ket{2}+  \ket{3})/2, \nonumber \\
                 &(\ket{0}+ i \ket{1}+ i \ket{2}-\ket{3})/2, \nonumber \\
                & (\ket{0}-i \ket{1}-i \ket{2}- \ket{3})/2 \}.              
\end{align}

\vspace{2cm}
Now, for calculating the sum of $6$ PCCs in the case of $d=5$ isotropic states and Werner states,
we consider the following choice of $6$ noncommuting bases:
\begin{widetext}
\begin{align}\label{ncb5}
 \{\ket{a_j}\}=&\{\ket{0}, \ket{1}, \ket{2}, \ket{3}, \ket{4}\} \nonumber \\
 \{\ket{b_j}\}=&\{(\ket{0}+ \omega^3 \ket{1}+  \omega \ket{2}+  \omega^4 \ket{3}+ \omega^2 \ket{4})/\sqrt{5},  \nonumber \\
                 &(\ket{0}+ \omega^4 \ket{1}+  \omega^3 \ket{2}+  \omega^2 \ket{3}+ \omega \ket{4})/\sqrt{5},  \nonumber \\
                 &(\ket{0}+  \ket{1}+\ket{2}+\ket{3}+\ket{4})/\sqrt{5}, \nonumber \\
                & (\ket{0}+ \omega \ket{1}+ \omega^2 \ket{2}+ \omega^3 \ket{3}+ \omega^4 \ket{4})/\sqrt{5}, \nonumber \\
                 &(\ket{0}+ \omega^2 \ket{1}+  \omega^4 \ket{2}+  \omega \ket{3}+ \omega^3 \ket{4})/\sqrt{5}\} \nonumber \\
  \{\ket{e_j}\}=&\{(\ket{0}+ e^{i\pi/5} \ket{1}+ e^{2i\pi/5} \ket{2}+e^{3i\pi/5} \ket{3}+e^{4i\pi/5}\ket{4})/\sqrt{5}, \nonumber \\
                & (\ket{0}+ \omega  e^{i\pi/5} \ket{1}+ \omega^2  e^{2i\pi/5} \ket{2}+ \omega^3  e^{3i\pi/5} \ket{3}
                + \omega^4  e^{4i\pi/5} \ket{4})/\sqrt{5}, \nonumber \\
                 &(\ket{0}+ \omega^2  e^{i\pi/5} \ket{1}+  \omega^4  e^{2i\pi/5} \ket{2}+  \omega  e^{3i\pi/5} \ket{3}+ \omega^3  e^{4i\pi/5} \ket{4})/\sqrt{5}, \nonumber \\
                 &(\ket{0}+ \omega^3  e^{i\pi/5} \ket{1}+  \omega  e^{2i\pi/5} \ket{2}+  \omega^4  e^{3i\pi/5} \ket{3}+ \omega^2  e^{i4\pi/5} \ket{4})/\sqrt{5},  \nonumber \\
                 &(\ket{0}+ \omega^4  e^{i\pi/5} \ket{1}+  \omega^3  e^{2i\pi/5} \ket{2}+  \omega^2   e^{3i\pi/5}\ket{3}+ \omega  e^{4i\pi/5} \ket{4})/\sqrt{5}\} \nonumber \\ 
   \{\ket{g_j}\}=&\{(\omega^2 \ket{0}+ \omega^3 \ket{1}+  \omega \ket{2}+  \omega^4 \ket{3}+  \ket{4})/\sqrt{5},  \nonumber \\
                 &(\omega \ket{0}+ \omega^4 \ket{1}+  \omega^3 \ket{2}+  \omega^2 \ket{3}+  \ket{4})/\sqrt{5},  \nonumber \\
                 &(\ket{0}+  \ket{1}+\ket{2}+\ket{3}+\ket{4})/\sqrt{5}, \nonumber \\
                 &(\omega^4\ket{0}+ \omega \ket{1}+ \omega^2 \ket{2}+ \omega^3 \ket{3}+  \ket{4})/\sqrt{5}, \nonumber \\
                 &(\omega^3 \ket{0}+ \omega^2 \ket{1}+  \omega^4 \ket{2}+  \omega \ket{3}+  \ket{4})/\sqrt{5}\} \nonumber \\
  \{\ket{k_j}\}=&\{(e^{4i\pi/5}\ket{0}+ e^{i\pi/5} \ket{1}+ e^{2i\pi/5} \ket{2}+e^{3i\pi/5} \ket{3}+\ket{4})/\sqrt{5}, \nonumber \\
                & (\omega^4  e^{4i\pi/5} \ket{0}+ \omega  e^{i\pi/5} \ket{1}+ \omega^2  e^{2i\pi/5} \ket{2}+ \omega^3  e^{3i\pi/5} \ket{3}
                +  \ket{4})/\sqrt{5}, \nonumber \\
                 &(\omega^3  e^{4i\pi/5} \ket{0}+ \omega^2  e^{i\pi/5} \ket{1}+  \omega^4  e^{2i\pi/5} \ket{2}+  \omega  e^{3i\pi/5} \ket{3}+  \ket{4})/\sqrt{5}, \nonumber \\
                 &(\omega^2  e^{i4\pi/5}\ket{0}+ \omega^3  e^{i\pi/5} \ket{1}+  \omega  e^{2i\pi/5} \ket{2}+  \omega^4  e^{3i\pi/5} \ket{3}+  \ket{4})/\sqrt{5},  \nonumber \\
                 &(\omega  e^{4i\pi/5}\ket{0}+ \omega^4  e^{i\pi/5} \ket{1}+  \omega^3  e^{2i\pi/5} \ket{2}+  \omega^2   e^{3i\pi/5}\ket{3}+  \ket{4})/\sqrt{5}\} \nonumber \\
 \{\ket{l_j}\}=&\{( \omega^3 \ket{0}+ \omega \ket{1}+  \omega^4 \ket{2}+   \omega^2  \ket{3}- \ket{4})/\sqrt{5},  \nonumber \\
                 &(\omega^4  \ket{0}+ \omega^3 \ket{1}+  \omega^2 \ket{2}+ \omega   \ket{3}-  \ket{4})/\sqrt{5}, \nonumber \\
                 &(\ket{0}+  \ket{1}+\ket{2}+\ket{3}-\ket{4})/\sqrt{5}, \nonumber \\
                & (\omega \ket{0}+ \omega^2 \ket{1}+   \omega^3 \ket{2}+ \omega^4 \ket{3}- \ket{4})/\sqrt{5}, \nonumber \\
                 &(\omega^2 \ket{0}+ \omega^4 \ket{1}+  \omega   \ket{2}+ \omega^3 \ket{3}-  \ket{4})/\sqrt{5}\},
\end{align}
\end{widetext}
where $\omega=2i\pi/5$.  It can be checked that the above
noncommuting bases are \textit{not} unbiased to each other.

\section{Derivation of Eq. (\ref{d3WPNS}) for the sum of four 
PCCs for the two-qutrit Werner-Popescu states}  \label{AWP}
For  $A_1=B_1=\sum_j a_j \ketbra{a_j}{a_j}$ in which the basis $\{\ket{a_j}\}$ is the computational basis and 
the eigenvalues $a_j$ are given by $a_0=+1$, $a_1=0$ and $a_2=-1$, the relevant single and joint
expectation values of the two-qutrit Werner-Popescu  states given by Eq. (\ref{WPd}) with $d=3$ are given by
\begin{align}
 \braket{A_1}&=\braket{B_1}=0,  \nonumber \\
 \braket{A^2_1}&=\braket{B^2_1}=\frac{2}{3},  \nonumber \\ 
 \braket{A_1B_1}&=\frac{2p}{3}. \nonumber
\end{align}
From the above expressions, it can be checked that the PCC in this case takes the value
\be
C_{A_1B_1}=p. \label{AWP1}
\ee

For  $A_2=B_2=\sum_j b_j \ketbra{b_j}{b_j}$, where the basis $\{\ket{b_j}\}$ is given in Eq. (\ref{csmub3})
and the eigenvalues $b_j$ are given by $b_0=0$, $b_1=\pm 1$ and $b_2=\mp 1$, the relevant  single and joint
expectation values are given by
\begin{align}
 \braket{A_2}&=\braket{B_2}=0, \nonumber \\
 \braket{A^2_2}&=\braket{B^2_2}=\frac{2}{3},  \nonumber \\ 
 \braket{A_2B_2}&=\frac{-2p}{3}. \nonumber
\end{align}
From the above expressions, it can be checked that the PCC  in this case is given by
\be
C_{A_2B_2}=-p. \label{AWP2}
\ee

For  $A_3=B_3=\sum_j e_j \ketbra{e_j}{e_j}$, where the basis $\{\ket{e_j}\}$ is given in Eq. (\ref{csmub3}) and 
the eigenvalues $e_j$ are given by $e_0=+1$, $e_1=0$ and $e_2=-1$, the relevant  single and joint
expectation values are given by
\begin{align}
 \braket{A_3}&=\braket{B_3}=0, \nonumber \\
 \braket{A^2_3}&=\braket{B^2_3}=\frac{2}{3}, \nonumber \\ 
 \braket{A_3B_3}&=\frac{-2p}{3}. \nonumber
\end{align}
From the above expressions, it can be checked that the PCC in this case takes the value
\be
C_{A_3B_3}=-p. \label{AWP3}
\ee

For  $A_4=B_4=\sum_j g_j \ketbra{g_j}{g_j}$, where the basis $\{\ket{g_j}\}$ is given in Eq. (\ref{csmub3})
and the eigenvalues $g_j$ are given by $g_0=+1$, $g_1=0$ and $g_2=-1$, the relevant single and joint
expectation values are given by
\begin{align}
 \braket{A_4}&=\braket{B_4}=0, \nonumber \\
 \braket{A^2_4}&=\braket{B^2_4}=\frac{2}{3}, \nonumber \\ 
 \braket{A_4B_4}&=\frac{-2p}{3}. \nonumber
\end{align}
From the above expressions, it can be checked that the PCC in this case is given by
\be
C_{A_4B_4}=-p. \label{AWP4}
\ee
Then  Eq. (\ref{d3WPNS}) follows from Eqs. (\ref{AWP1})-(\ref{AWP4}).

\section{Two MUBs used for checking whether the sum of two PCCs for the two-qudit pure states in $d=4$ is linearly related with Negativity}\label{LRNd=4}
 We consider the sum of two PCCs $|C_{A_1B_1}|+|C_{A_2B_2}|$ for the two-qudit pure states in $d=4$ given by Eq. (\ref{q4Sc}) with respect to the following choice of 
observables: $A_1=B_1=\sum^3_{j=0} a_j \ketbra{a_j}{a_j}$ and $A_2=B_2=\sum^3_{j=0} b_j \ketbra{b_j}{b_j}$, where $\{\ket{a_j}\}$ is the computational basis
with the eigenvalues $a_0=+2$, $a_1=+1$, $a_2=-1$ and $a_3=-2$
and $\{\ket{b_j}\}$ is a three-parameter family of basis which is mutually unbiased to the computational basis given by
\ba
\ket{b_j}&=&\ket{0}+\frac{e^{i(2\pi/d)j}}{\sqrt{d}}(e^{i\phi_x})\ket{1}+\frac{e^{i(4\pi/d)j}}{\sqrt{d}}(e^{i2\phi_y})\ket{2} \nonumber \\
&+&\frac{e^{i(6\pi/d)j}}{\sqrt{d}}(e^{i3\phi_z})\ket{3},
\ea
with $0 \le \phi_x, \phi_y, \phi_z \le 2\pi$ and the eigenvalues $b_0=+2$, $b_1=+1$, $b_2=-1$ and $b_2=-2$. 
We have numerically checked whether there exists any choice of above such two MUBs for which the above sum of
two PCCs has a linear relationship with Negativity by varying over all choices of  two MUBs with respect to the parameters   $\phi_x$, $\phi_y$ and  $\phi_z$.
From this numerical search, it has been found that there does not exist any such two MUBs for which the sum of two PCCs is linearly related with Negativity.
\end{document}